\documentclass[traditabstract]{aa} 
\usepackage{graphicx}
\usepackage{txfonts}
\usepackage[ansinew]{inputenc}
\usepackage{mathrsfs}
\usepackage{textcomp,booktabs}
\usepackage{graphicx}
\usepackage[hang]{subfigure}
\usepackage{multirow}
\usepackage{natbib}
\usepackage{lscape}
\usepackage{longtable}

\bibpunct{(}{)}{;}{a}{}{,} 
\newcommand{\Diff}[2]{\frac{\mathrm{d} #1}{\mathrm{d} #2}}

\newcommand{\Difft}[2]{\mathrm{d} #1 \slash \mathrm{d} #2}

\newcommand{\unit}[1]{~\mathrm{#1}}

\newcommand{\fract}[2]{{#1}\slash{#2}}

\newcommand{\fermilat}{\emph{Fermi}-LAT}
\newcommand{\gray}{$\gamma$-ray}
\newcommand{\grays}{$\gamma$-rays}
\newcommand{\fermi}{\emph{VHE-HEIndex}~}
\newcommand{\intg}{\emph{IntVHELumi}~}
\newcommand{\pileup}{\emph{PileUp}~}
\newcommand{\conca}{\emph{VHE\-Concavity}~}
\newcommand{\murm}{%
  \ifmmode
    \mathchoice
        {\hbox{\normalsize\textmu}}
        {\hbox{\normalsize\textmu}}
        {\hbox{\scriptsize\textmu}}
        {\hbox{\tiny\textmu}}%
  \else
    \textmu
  \fi
}

\begin{document}
   \title{Limits on the extragalactic background light 
in the \textit{Fermi} era}

   \author{Manuel Meyer
          \inst{1}\thanks{\email{manuel.meyer@physik.uni-hamburg.de}}
          \and
          Martin Raue
	  \inst{1}
	  \and
	  Daniel Mazin
	  \inst{2}
	  \and
	  Dieter Horns
	  \inst{1}
          }

   \institute{Institut f\"ur Experimentalphysik, University of Hamburg,
              Luruper Chaussee 149, 22761 Hamburg, Germany
         \and
             Institut de Fisica d'Altes Energies (IFAE),\\
	     Edifici Cn. Universitat Autonoma de Barcelona, 08193 Bellaterra, Spain\\
             }

   \date{Accepted for publication in Astronomy \& Astrophysics}

  \abstract
{
Very high energy (VHE, energy $E \gtrsim 100$\,GeV) \grays~from cosmological sources are attenuated 
due to the interaction with photons of the extragalactic background light (EBL) in the ultraviolet to infrared wavelength band. 
The EBL, thus, leaves an imprint on the observed energy spectra of these objects. 
In the last four years, the number of extragalactic VHE sources discovered with imaging atmospheric Cherenkov telescopes (IACTs),
such as MAGIC, H.E.S.S., and VERITAS, has doubled.
Furthermore, the measurements of the \emph{Fermi} satellite brought new insights into the intrinsic spectra of the sources at GeV energies.
In this paper,
upper limits on the EBL intensity are derived by considering the most extensive VHE source sample ever used in this context.
This is accomplished by constructing a large number of generic EBL shapes and
combining spectral informations from \emph{Fermi} and IACTs together with minimal assumptions about the source physics
at high and very high \gray~energies.
The evolution of the EBL with redshift is accounted for and the possibility of the formation of an electromagnetic cascade
and the implications on the upper limits are explored.
The EBL density at $z=0$ is constrained over a broad wavelength range between 0.4 and 100\,\murm m. 
At optical wavelengths, the EBL density is constrained below 24\,nW\,m$^{-2}$\,sr$^{-1}$ and below 5\,nW\,m$^{-2}$\,sr$^{-1}$ between 8\,\murm m and 31\,\murm m.
}

   \keywords{cosmology: diffuse radiation --
               galaxies: BL Lacertae objects: general  --
               galaxies: active --
	       infrared: diffuse background
               }

   \maketitle
%

\section{Introduction}
The extragalactic background light (EBL) is the diffuse and isotropic radiation field from ultraviolet to far infrared wavelengths 
\citep[see e.g.][for a review]{2001ARA&A..39..249H,2005PhR...409..361K}.
It originates from the starlight integrated over all epochs and starlight emission reprocessed by interstellar dust.
These two distinct components lead to two maxima in the spectral energy distribution (SED) of the EBL, the first at $\sim$ 1\,\murm m (starlight) and the second at 
$\sim$ 100\,\murm m (dust).
Narrow spectral features like absorption lines are smeared out in the integration over redshift leading to a smooth shape of the EBL at $z=0$.
Further contributions may come from diffuse emission from galaxy clusters \citep{2007ApJ...671L..97C}, unresolved active galactic nuclei \citep{2006A&A...451..443M}, 
Population III stars \citep[e.g.][]{2002MNRAS.336.1082S,2009A&A...498...25R},
or exotic sources like dark matter powered stars in the early universe \citep{MaurerDS}.
The SED of the EBL at $z=0$ comprises information about the star and galaxy formation rates and the dust content of galaxies.
Direct measurements of the EBL are, however, impeded, especially in the infrared, by foreground emission such as the zodiacal light \citep{1998ApJ...508...25H}.
Therefore, upper and lower limits are often the only available information about the EBL density.
Lower limits are derived from integrated galaxy number counts e.g. by the \textit{Hubble Space Telescope} in the optical \citep{2000MNRAS.312L...9M} and the \textit{Spitzer} telescope in the infrared \citep{2004ApJS..154...39F}.
Several authors have modeled the EBL in the past \citep[e.g.][]{2005AIPC..745...23P,2006ApJ...648..774S,2008A&A...487..837F,2010A&A...515A..19K,2011MNRAS.410.2556D}.
Although the approaches forecast different EBL densities at $z=0$, the most recent models more or less agree on the overall EBL shape.

The observations of very high energy (energy $E \gtrsim 100$\,GeV; VHE) \grays~from extragalactic sources with imaging atmospheric Cherenkov telescopes (IACTs) has opened a new window to constrain the EBL density.
Most of the extragalactic $\gamma$-ray sources are active galactic nuclei (AGN), especially blazars \citep[see e.g.][]{1995PASP..107..803U}.
The \grays~from these objects are attenuated by the pair production mechanism:
$\gamma_\mathrm{VHE} + \gamma_\mathrm{EBL} \rightarrow e^+ + e^-$ \citep{1962Nikishov,1966PhRvL..16..479J,1967PhRv..155.1404G}.
If assumptions are made about the properties of the intrinsic blazar spectrum, a comparison with the observed spectrum allows to place upper limits on the EBL intensity
 \citep[e.g.][]{1992ApJ...390L..49S}.
In this context, the spectra of Markarian (Mkn) 501 during an extraordinary flare
\citep{1999A&A...349...11A} and of the distant blazar H\,1426+482 \citep{2003A&A...403..523A}
resulted in the first constraints of the EBL density from mid to far infrared (MIR and FIR) wavelengths. 
With the new generation of IACTs, limits were derived from the spectra of
1ES\,1101-232 and H\,2356-309 \citep{2006Natur.440.1018A} and 1ES\,0229+200 \citep{2007A&A...475L...9A} in the near infrared (NIR), 
and in the optical from the MAGIC observation of 3C\,279 \citep{2008Sci...320.1752M}.
\citet[][henceforth MR07]{2007A&A...471..439M} use a sample of all at that time known blazars 
and test a large number of different EBL shapes to derive robust constraints over a large wavelength range.
The authors exclude EBL densities that produce VHE spectra, characterized by $\Difft{N}{E} \propto E^{-\Gamma}$,
with $\Gamma < \Gamma_\mathrm{limit}$ (being $\Gamma_\mathrm{limit} = 1.5$ for realistic and $\Gamma_\mathrm{limit} = 2/3$ for extreme scenarios) 
or an exponential pile up at highest energies.

With the advent of the large area telescope (LAT) on board the \emph{Fermi} satellite \citep{2009ApJ...697.1071A} and its unprecedented sensitivity at high energies 
($100\unit{MeV} \lesssim E \lesssim 100\unit{GeV}$; HE), further possibilities arose to confine the EBL density.
Bounds can be derived either by considering solely \fermilat~observations of AGN and gamma-ray bursts \citep{2010ApJ...723.1082A,2010A&A...520A..34R} or by combining HE with VHE spectra \citep[e.g.][]{2010ApJ...714L.157G,2011ApJ...733...77OO}.
It has also been proposed that the \fermilat~can, in principle, measure the EBL photons upscattered by electrons in lobes of radio galaxies directly \citep{2008ApJ...686L...5G}.
Attenuation limits can also be estimated by modeling the entire blazar SED in order to forecast the intrinsic VHE emission 
\citep{2002MNRAS.336..721K,2010ApJ...715L..16M}.

In this paper, results from the recently published \emph{Fermi} two year catalog \citep[][henceforth 2FGL]{2FGL} 
together with a comprehensive VHE spectra sample are used to place upper limits on the EBL density.
This approach relies on minimal assumptions about the intrinsic spectra.
The VHE sample is composed of spectra measured with different instruments,
thereby ensuring that the results are not influenced by the possible systematic bias of an individual instrument or observation. 

The article is organized as follows. In Section \ref{sec:grid} the calculation of the attenuation is presented in order to correct the observed spectra for absorption.
The resulting intrinsic spectra are subsequently described with analytical functions.
Section \ref{sec:excl} outlines in detail the different approaches to constrain the EBL before
the selection of VHE spectra is addressed in Section \ref{sec:samples}. The combination of VHE and HE spectra of variable sources will also be discussed.
The results are presented in Section \ref{sec:results} before concluding in Section \ref{sec:concl}.
Throughout this work a standard $\Lambda$CDM cosmology is assumed with $\Omega_m = 0.3$, $\Omega_\Lambda = 0.7$, and $h=0.72$.

\section{Intrinsic VHE gamma-ray spectra}
\label{sec:grid}
 The intrinsic energy spectrum $\Difft{N_\mathrm{int}}{E}$ of a source at redshift $z_0$ at the measured energy $E$ differs from the observed spectrum 
$\Difft{N_\mathrm{obs}}{E}$ due to the interaction of source photons with the photons of the EBL which is most commonly expressed as 
\begin{equation}
 \Diff{N_\mathrm{obs}}{E} = \Diff{N_\mathrm{int}}{E} \times \exp\left[-\tau_\gamma(E,z_0)\right]\label{eqn:abs}.
\end{equation}
The strength of the attenuation at energy $E$ is given by the optical depth $\tau_\gamma$: 
a threefold integral over the distance $\ell$, the cosine $\mu$ of the angle between the photon momenta, and the energy $\epsilon$ of the EBL photons \citep[e.g.][]{2005ApJ...618..657D},
\begin{eqnarray}
\tau_{\gamma}(E,z_0) &=&\nonumber\\
  &{}&\hskip-30pt 
\int\limits_0^{z_0} \mathrm d \ell(z) \int\limits_{-1}^{+1} \mathrm d \mu \frac{1 - \mu}{2}  
\int\limits_{\epsilon^\prime_\mathrm{thr}}^{\infty}\mathrm d \epsilon^\prime
n_\mathrm{EBL}(\epsilon^\prime, z)
\sigma_{\gamma\gamma}(E^\prime,\epsilon^\prime,\mu).
\label{eqn:tau}
\end{eqnarray}
The primed values correspond to the redshifted energies and $n_\mathrm{EBL}(\epsilon^\prime, z)$ denotes the comoving EBL photon number density.
The threshold energy for pair production is given by $\epsilon_\mathrm{thr}^\prime = \epsilon_\mathrm{thr}(E^\prime,\mu)$ with $E^\prime = E(1+z)$.
The cross section for pair production, $\sigma_{\gamma\gamma}$, is strongly peaked at a wavelength \citep[e.g.][]{2000A&A...359..419G}
\begin{equation}
\lambda_\ast = \frac{hc}{\epsilon_\ast} \approx  1.24 \left(\frac{E}{\mathrm{TeV}}\right)\,\murm\mathrm{m}\label{eqn:ppwave},
\end{equation}
and, therefore, VHE \grays~predominantly interact with EBL photons from optical to FIR wavelengths.

The comoving EBL photon density is described 
here by splines constructed from a grid in ($\lambda{,}\nu I_\nu$)-plane (see MR07 for further details).
This ensures independence of EBL model assumptions and allows for a great variety of EBL shapes to be tested.
Furthermore, the usage of splines drastically reduces the effort to compute the complete
threefold integral of Eq. \ref{eqn:tau} numerically as shown in MR07. 

\begin{figure}[tb]
\centering
 \includegraphics[angle= 270, width = 1. \linewidth]{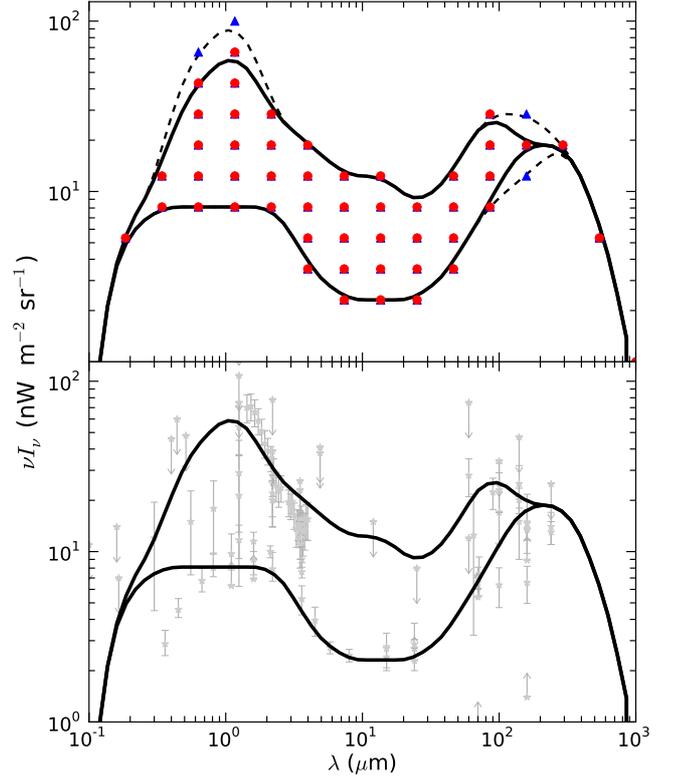}
\caption{
Upper panel: Grid in wavelength versus the energy density 
of the EBL used to construct the EBL shapes for testing (red bullets).
Also shown are the minimum and maximum shape tested (solid lines) and the same for the grid of MR07 (blue triangles; dashed lines).
Lower panel: Minimum and maximum EBL shape tested versus EBL limits and measurements
\citep[light gray symbols, see][and references therein]{2011arXiv1106.4384R}.
}
\label{fig:grid-basic}
\end{figure}

Each spline is defined by the choice of knot points and weights from the grid in the $(\lambda{,}\nu I_\nu)$-plane.
The grid is bound by a minimum and a maximum shape, shown in Figure \ref{fig:grid-basic}.
The setup of grid points is taken from MR07.
The minimum shape tested is set to reproduce the lower limits from the galaxy number counts from \emph{Spitzer} \citep{2004ApJS..154...39F} 
while the maximum shape roughly follows the upper limits derived from measurements.
To reduce the computational costs, the extreme cases considered by MR07 of the EBL density in the optical and near infrared (NIR) are not tested. 
Moreover, with current VHE spectra the EBL intensity is only testable up to a wavelength $\lambda \approx 100$\,\murm m so no additional grid points beyond this wavelength are used.
In total this range of knots and weights allows for 1,920,000 different EBL shapes.
A much smaller spacing of the grid points is not meaningful
as small structures are smeared out in the calculation of $\tau_\gamma$ and the EBL can be understood as a superposition of black bodies
that are not arbitrarily narrow in wavelength \citep[MR07,][]{RauePhD}.

In most previous studies, no EBL evolution with redshift is assumed when computing EBL upper limits using VHE $\gamma$-ray observations.
Neglecting the evolution leads to an overestimation of the optical depth between 10\,\% ($z = 0.2$) and 35\,\% ($z= 0.5$) \citep[][]{2008IJMPD..17.1515R}
and, consequently, too rigid upper limits (see Appendix \ref{sec:evo}).
In this study, the $z$ evolution is accounted for by a phenomenological ansatz \citep[e.g.][]{2008IJMPD..17.1515R}:
the effective cosmological photon number density scaling is changed from $n_\mathrm{EBL} \propto (1 + z)^3$ to $n_\mathrm{EBL} \propto (1 + z)^{3-f_\mathrm{evo}}$.
 For a value of $f_\mathrm{evo}= 1.2$ a good agreement is found between this simplified approach and complete EBL model calculations for redshifts $z \lesssim 0.7$  
\citep{2008IJMPD..17.1515R}.
Including the redshift evolution of the EBL in general decreases the attenuation compared to the no-evolution case and, 
therefore, weaker EBL limits are expected 

The intrinsic $\gamma$-ray spectrum for a given EBL shape and measured $\gamma$-ray spectrum 
is reconstructed by solving Eq. \ref{eqn:abs} for $\Difft{N_\mathrm{int}}{E}$.
For a spectrum with $n$ energy bins the relation reads
\begin{equation}
 \left(\Diff{N_\mathrm{int}}{E}\right)_i = \left(\Diff{N_\mathrm{obs}}{E}\right)_i \times \exp\left[{\tau_\gamma(E_i,z_0)}\label{eqn:deabs}\right], \quad i = 1,\ldots,n,
\end{equation}
where the energy of the logarithmic bin center is denoted by $E_i$. 
A systematic error is introduced by using $\tau_\gamma$ calculated for the energy at  
the bin center since, on the one hand, the attenuation can change  
dramatically within relatively wide energy bins and, on the other  
hand, the mean attenuation actually depends on the intrinsic spectral  
shape in the energy bin \citep{2009ApJ...691L..91S}. 
The introduced error is studied by
comparing $\tau_\gamma$ with an averaged value of the optical depth over the highest energy bin for the spectra that are attenuated most.
These spectra are described with an analytical function $f(E)$ (a power or broken power law, cf. Table \ref{tab:models}) and the averaged
optical depth $\langle \tau_\gamma \rangle$ is found to be
\begin{equation}
 \langle \tau_\gamma \rangle = \frac{\int\limits_{\Delta E}\tau_\gamma(E,z) f(E)\,\mathrm{d} E}{\int\limits_{\Delta E} f(E)\,\mathrm{d}E}.
\end{equation}
The results are summarized in Table \ref{tab:diffave}. The ratios $\langle\tau_\gamma\rangle/\tau_\gamma$ are close to, but always smaller than, one
 and the optical depth is overestimated by $<$ 5\,\%.  Thus, the simplified approach adds marginally to the uncertainties of the upper limits.

\begin{table}[htb]
\centering
\caption{
Comparison between the optical depth at the logarithmic bin center of the highest energy bin and the averaged value over the bin width.}
\label{tab:diffave}
\begin{scriptsize}
\begin{tabular}{c|ccc|ccc}
\hline
\hline
\multirow{2}{*}{Source} & \multicolumn{3}{c|}{Minimum EBL shape tested} & \multicolumn{3}{c}{Maximum EBL shape tested} \\
{} & $\tau_\gamma$ & $\langle\tau_\gamma\rangle$ &
$\fract{\langle\tau_\gamma\rangle}{\tau_\gamma}$ & 
$\tau_\gamma$ & $\langle\tau_\gamma\rangle$ &
$\fract{\langle\tau_\gamma\rangle}{\tau_\gamma}$ \\
\hline
3C\,279 		& 3.48 & 3.34 & 0.96 & 18.33 & 17.61 & 0.96 \\
H\,1426+428 	& 2.54 & 2.53 & 0.99 & 12.61 & 12.48 & 0.99 \\
1ES\,1101-232 	& 2.69 & 2.68 & 1.00 & 13.62 & 13.57 & 1.00 \\ 
Mkn\,501 	& 3.27 & 3.21 & 0.98 & 11.86 & 11.67 & 0.98 \\
\hline 
\end{tabular}
\end{scriptsize}
\tablefoot{
The HEGRA spectrum is used here for Mkn\,501, see Table \ref{tab:samples} for the references.
}
 
\end{table}

The intrinsic spectra obtained by means of Eq. \ref{eqn:deabs} will be described by analytical functions in order to test the fit parameters for their physical feasibility.
An analytical description of the spectrum is determined by fitting a series of functions listed in Table \ref{tab:models}. 
A $\chi^2$-minimization algorithm \citep[utilizing the MINUIT package routines, see][]{Minuit} is employed,
 starting with the first function of the table, a simple power law. 
The fit is not considered valid if the corresponding probability is $P_\mathrm{fit}(\chi^2) < 0.05$. 
In this case the next function with more model parameters from Table \ref{tab:models} is evaluated.
For a given energy spectrum of $n$ data points, only functions are  
examined with $n-1 > 0$ degrees of freedom.
If more than one fit results in an acceptable fit probability, an $F$-Test is used to determine the preferred hypothesis. 
The parameters of the model with more fit parameters are examined if the test results in a 95\,\% probability that description of the data has improved.

\begin{table*}[htb]
 \caption{Analytical functions fitted to the deabsorbed spectra.}
 \label{tab:models}
\centering
 \begin{tabular}{lclc}
  \hline
  \hline
  Description & Abbreviation & Formula $\Difft{N_\mathrm{int}}{E}$& \# of parameters\\
\hline
  Power law & \multirow{2}{*}{PL} & \multirow{2}{*}{$\mathit{PL}_1$} & \multirow{2}{*}{2}\\
  {} & {} & {} & {}\\
  Broken power law & \multirow{2}{*}{BPL} & \multirow{2}{*}{$\mathit{PL}_1 \times \mathit{CPL}_{12}$} &  \multirow{2}{*}{4}\\
  with transition region& {} & {} & {} \\
  {} & {} & {} & {}\\
  Broken power law with transition region & \multirow{2}{*}{SEBPL} &  
    \multirow{2}{*}{
	$\mathit{PL}_1\times \mathit{CPL}_{12} \times \mathit{Pile}$ } &  \multirow{2}{*}{6}\\
 and super-exponential pile up & {} & {} & {}\\
  {} & {} & {} & {}\\
  Double broken power law & \multirow{2}{*}{DBPL} & \multirow{2}{*}{$\mathit{PL}_1\times \mathit{CPL}_{12} \times \mathit{CPL}_{23}$} &  \multirow{2}{*}{6}\\
  with transition region& {} & {} & {} \\
  {} & {} & {} & {}\\
  Double broken power law & \multirow{2}{*}{SEDBPL} & \multirow{2}{*}{
$\mathit{PL}_1\times \mathit{CPL}_{12} \times \mathit{CPL}_{23} \times \mathit{Pile}$} &  \multirow{2}{*}{8}\\
  with super-exponential pile up& {} & {} & {} \\
  {} & {} & {} & {}\\
\hline
 \end{tabular}
\tablefoot{The functions are a power law, a curved power law and a super exponential pile up defined as 
$\mathit{PL}_i = N_0\,E^{-\Gamma_i},~
\mathit{CPL}_{ij} = \left[ 1 + \left(\fract{E}{E^\mathrm{break}_i}\right)^{f_i} \right]^{\fract{(\Gamma_i - \Gamma_j)}{f_i}},~\mathrm{and}~
\mathit{Pile} = \exp\left[\left(E / E_\mathrm{pile}\right)^\beta\right]$, respectively,
where all energies are normalized to 1\,TeV.
The smoothness parameters $f_i$ are held constant and the break energies $E^\mathrm{break}_i$ are forced to be positive.
Only positive pile up, i.e. $E_\mathrm{pile} > 0$, is tested.
}
\end{table*}

\section{Exclusion criteria for the EBL shapes}
\label{sec:excl}
In the following, arguments to exclude EBL shapes will be presented.
While the first criteria are based on the expected concavity of the intrinsic VHE spectra,
the second set of criteria relies on the integral of the intrinsic VHE emission. 

\subsection{Concavity}
\label{sec:curvature}
Observations have led to the commonly accepted picture that particles are accelerated in jets of AGN thereby producing non-thermal radiation.
The SED of these objects is dominated by two components.
The first low-energy component from infrared to X-ray energies is due to synchrotron radiation from a distribution of relativistic electrons. 
The second component responsible for HE and VHE emission can be explained by several different emission models.
In leptonic blazar models, photons are upscattered by the inverse Compton (IC) process.
The involved photon fields originate from synchrotron emission \citep[e.g.][]{1996ApJ...461..657B},
the accretion disk \citep{1993ApJ...416..458D}, or the broad line region \citep[e.g.][]{1994ApJ...421..153S}.
In hadronic blazar models, on the other hand, $\gamma$-ray emission is produced either by proton synchrotron radiation \citep[e.g.][]{2001APh....15..121M} or photon pion production \citep[e.g.][]{2003APh....18..593M}.
These simple emission models which commonly describe the measured data satisfactorily, do not predict
a spectral hardening in the transition from HE to VHE nor within the VHE band.
This is also confirmed by observations of nearby sources.
On the contrary, the spectral slope is thought to become softer with energy, either due to Klein-Nishina effects in leptonic scenarios and / or 
a cut off in the spectrum of accelerated particles.

However, in more specific scenarios a spectral hardening is possible. 
If mechanisms like, e.g., second order IC scattering, internal photon absorption \citep[e.g.][]{2008MNRAS.387.1206A},
comptonization of low frequency radiation by a cold ultra-relativistic wind of particles \citep{2002A&A...384..834A},
or multiple HE and VHE \gray~emitting regions in the source \citep{2011arXiv1108.4568L}
contribute significantly to the overall spectrum, convex curvature or an exponential pile can indeed occur.
Nevertheless, neither of these features has been observed with certainty in nearby sources. 
Furthermore, it would imply serious fine tuning 
if such components appeared in all examined sources in the transition from the optical thin, i.e. $\tau_\gamma < 1$, to optical thick regime, $\tau_\gamma \ge 1$. 
This seems unlikely, considering the large number of EBL shapes tested.
Consequently, EBL shapes leading to an intrinsic VHE spectrum which is not concave will be excluded.
This expectation is formulated through three test criteria:

\paragraph{(i) \fermilat~spectrum as an upper limit.}
With the launch of the \emph{Fermi} satellite and the 
current generation of IACTs,
there is an increasing number
of broad-band AGN energy spectra measured
in the HE and VHE domains.
Thus, the least model dependent approach is to test spectra against a convex curvature in the transition from HE to VHE by
regarding the spectral index measured by \emph{Fermi}, $\Gamma_\mathrm{HE}$, as a limit on the reconstructed intrinsic index at VHE, $\Gamma$.
Hence, the intrinsic VHE spectrum is regarded as unphysical if the following condition is met,
\begin{equation}
\Gamma + \sigma_{\mathrm{stat}} + \sigma_\mathrm{sys} < \Gamma_\mathrm{HE} - \sigma_\mathrm{HE,~stat}.
\end{equation}
The statistical error $\sigma_{\mathrm{stat}}$ is estimated from the fit of an analytical function to the intrinsic spectrum 
whereas the systematic uncertainty $\sigma_\mathrm{sys}$ is used that is estimated by the respective instrumental team.
The statistical
uncertainty $\sigma_{\mathrm{HE,~stat}}$ of the \fermilat~spectral index is given by the 2FGL or the corresponding publication, see Section \ref{sec:samples}.
This exclusion criterion will be referred to as \fermi criterion in the following. 
Note that this is \emph{not} the same criterion as used by \citet{2011ApJ...733...77OO}. 
They assume that the VHE index should be equal to the index measured with the \fermilat. 

\paragraph{(ii) Super exponential pile up.}
Furthermore, shapes will be excluded that lead to an intrinsic VHE spectrum that piles up super exponentially at highest energies.
This is the case if it is best described by the analytical functions abbreviated SEBPL or SEDBPL, see Table \ref{tab:models}, and 
the pile-up energy is positive within a $1\sigma$ confidence, 
\begin{equation}
 E_\mathrm{pile} - \sigma_\mathrm{pile} > 0.\label{eqn:pileup}
\end{equation}
This additional independent exclusion criterion 
relies solely on VHE observations which are subject to the attenuation in contrast to \fermilat~observations.
It will be denoted as \pileup criterion throughout this study.

\paragraph{(iii) VHE concavity.}
In the case that the intrinsic spectrum is best described by either a BPL or a DBPL, 
it is considered as convex if the following inequalities are \textit{not} fulfilled,
\begin{eqnarray}
\Gamma_1 - \sigma_1 &\leqslant& \Gamma_2 + \sigma_2 \nonumber\\
{and}\quad\Gamma_2 - \sigma_2 &\leqslant& \Gamma_3 + \sigma_3 ~\mathrm{(DBPL)},\label{eqn:convex}
\end{eqnarray}
and the corresponding EBL shape will be rejected.
Again, 1$\sigma$ uncertainties of the fitting procedure are used. This criterion will be referred to as \textit{VHEConcavity}.
It is very similar to the argument formulated in (ii) as intrinsic spectra that show an exponential rise may often be equally well described by a BPL or DBPL.
However, with the \conca criterion also intrinsic spectra can be excluded that show only mild convexity,
i.e. no exponential pile up. 

\subsection{Cascade emission and energy budget}
\label{sec:integral}
In this Section two new approaches are introduced that are based on the integrated intrinsic emission.
These methods rely on a number of parameters, whose values are, so far, not accurately determined by observations or for which only upper and lower limits exist.
Therefore, the following two criteria have to be regarded as a theoretical motivated possibility to constrain the EBL in the future. 
As it will be shown in Section \ref{sec:results}, the final upper limits are not improved by these criteria and are, thus, independent of the model parameters chosen here.

\subsubsection{Cascade emission}
\label{sec:cascade}
EBL photons that interact with VHE \grays~produce $e^+e^-$ pairs. 
These secondary pairs can generate HE radiation by upscattering cosmic microwave background (CMB) photons by means of the IC process.
This initiates an electromagnetic cascade as these photons can again undergo pair production 
\citep[e.g.][]{1987MNRAS.227..403S,1994ApJ...423L...5A,2002ApJ...580L...7D,2009ApJ...703.1078D,2011arXiv1106.5508K}.
The amount of cascade radiation, that points back to the source, depends on the field strength $B_\mathrm{IGMF}$ of the intergalactic magnetic field and its correlation length $\lambda_B$.
The values of $B_\mathrm{IGMF}$ and $\lambda_B$ are unknown and only upper and lower limits exist \citep[see e.g.][for a compilation of limits]{2009PhRvD..80l3012N}.
If the field strength is large (see Eq. \ref{eqn:defl}) or if the correlation length
is small compared to the cooling length $ct_\mathrm{cool}$ of the $e^+e^-$ pairs for IC scattering,
the pairs are quickly isotropized and extended halos of $\gamma$-ray emission form around the initial source \citep[e.g][]{1994ApJ...423L...5A,2002ApJ...580L...7D,2009PhRvD..80b3010E,2009ApJ...703.1078D}.
Furthermore, the time delay of the cascade emission compared to the primary emission depends on $B_\mathrm{IGMF}$ and $\lambda_B$. 
VHE \grays~need to be produced for a sufficiently long period so that the reprocessed radiation is observable \citep[e.g.][]{2011ApJ...733L..21D}. 

The cascade emission has been used to place lower limits on $B_\mathrm{IGMF}$ and $\lambda_B$ 
by assuming a certain EBL model \citep{2010Sci...328...73N,2010MNRAS.406L..70T,2011MNRAS.tmp..570T,2011ApJ...733L..21D,2011ApJ...727L...4D,2011ApJ...735L..28H}.
Conversely, one can place upper limits on the EBL density under the assumption of a certain magnetic field strength.
This novel approach is followed here whereas, in previous studies, the cascade emission is neglected when deriving upper limits on the EBL density.
A higher EBL density leads to a higher production of $e^+e^-$ pairs and thus to a higher cascade emission that is potentially detectable with the \fermilat.
If the predicted cascade radiation exceeds the observations of the \fermilat, the corresponding EBL shape can be excluded.
Conservative upper limits are derived if the following assumptions are made:
(i) The HE emission of the source is entirely due to the cascade. 
(ii) The observed VHE spectrum is fitted with a power law with a super exponential cut off at the highest measured energy of the spectrum. 
This minimizes the reprocessed emission and allows to consider only the first generation of the cascade.
(iii) The $e^+e^-$ pairs are isotropized in the intergalactic magnetic field, minimizing the reprocessed emission.  
This condition is equal to the demand that the deflection angle $\vartheta$ of the particles in the magnetic field is $\approx \pi$.
 Assuming $\lambda_B \gg ct_\mathrm{cool}$, the deflection angle for electrons with an energy $\gamma mc^2 \approx E / 2$, where $E$ is the energy of the primary $\gamma$ ray,
can be approximated by \citep{2010MNRAS.406L..70T,2010Sci...328...73N}
\begin{equation}
\vartheta\approx \frac{ct_\mathrm{cool}}{R_\mathrm{L}} = 1.17\left(\frac{B_\mathrm{IGMF}}{10^{-15}\unit{G}}\right)(1+z_r)^{-4}\left(\frac{\gamma}{10^6}\right)^{-2},
\label{eqn:defl}
\end{equation}
with $z_r$ the redshift where the IC scattering occurs and
$R_\mathrm{L}$ the Larmor radius.
The IC scattered $e^+e^-$ pairs give rise to \grays~with energy $\epsilon \approx \gamma^2h\nu_\mathrm{CMB} \approx 0.63 (E/\mathrm{TeV})^2\,\mathrm{GeV}$,
with $h\nu_\mathrm{CMB} = 634$\,\murm eV the peak energy of the CMB.
The $\gamma$ factor in Eq. \ref{eqn:defl} can be eliminated in favor of $\epsilon$,
and, solving for $B_\mathrm{IGMF}$, the pairs are isotropized if $B_\mathrm{IGMF} \approx 4.2\times10^{-15}~(1+z_r)^4(\epsilon /\mathrm{GeV}) \unit{G} \approx 5 \times 10^{-13} \unit{G}$
for $\epsilon = 100$\,GeV, the maximum energy measured with the \fermilat~considered here 
and the maximum redshift where the IC scattering can occur, i.e. the redshift of the source.\footnote{
Accordingly, this $B$-field value ensures isotropy regardless were the IC scattering occurs.}
This value of $B_\mathrm{IGMF}$ is in accordance with all experimental bounds \citep[see e.g.][especially Figures 1 and 2]{2009PhRvD..80l3012N} 
For correlation lengths $\lambda_B \gg ct_\mathrm{cool} \approx 0.65 (E/\mathrm{TeV})^{-1}(1+z_r)^{-4}$\,Mpc $\approx \mathcal{O}(\mathrm{Mpc})$ 
the most stringent constraints come from Faraday rotation measurements \citep{1976Natur.263..653K,1999ApJ...514L..79B}
which limit $B_\mathrm{IGMF} \lesssim 10^{-9}$~G. 
Furthermore, the adopted value cannot be excluded neither with possible observations of deflections of ultra-high energy cosmic rays \citep{1995ApJ...455L..21L}
nor with constrained simulations of magnetic fields in galaxy clusters, both setting an upper limit on $B_\mathrm{IGMF} \lesssim 10^{-12}$\,G \citep{2005JCAP...01..009D,2009MNRAS.392.1008D}.

For value of $B_\mathrm{IGMF}$, the cascade emission is detectable if a steady \gray~emission of the source for the last $\Delta t \approx 10^6$ years is assumed
\citep[see e.g.][]{2011ApJ...733L..21D,2011MNRAS.tmp..570T}.
Other energy loss channels apart from IC scattering like synchrotron radiation or plasma instabilities \citep[][]{2011arXiv1106.5494B} are neglected.
However, if the latter are present, the field strength is even higher, or the lifetime of the VHE source is shorter, 
no significant cascade emission is produced or it has not reached earth so far.

\begin{figure}[htb]
 \centering
 \includegraphics[width = .8\linewidth, angle = 270]{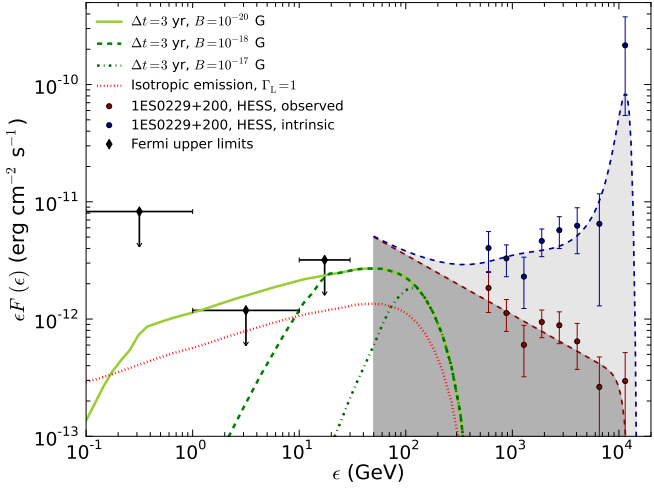}
\caption{Cascade emission for a certain EBL shape and the VHE spectrum of 1ES\,0229+200 \citep{2007A&A...475L...9A}. 
The observed spectrum (dark red points and line) is fitted with a power law with an exponential cut off and corrected for the EBL absorption (dark blue dashed line and points).
The green lines show the cascade emission resulting from the reprocessed flux (light gray shaded area) for a constant emission over the last three years and different magnetic field strengths. 
The red dotted line shows the reprocessed emission if the $e^+e^-$ pairs are isotropized. The latter does not overproduce the \emph{Fermi} upper limits \citep[black diamonds][]{2010MNRAS.406L..70T} 
and hence the corresponding EBL shape is not excluded.
The light and dark gray area together are equal to the integrated flux that is compared to the Eddington luminosity (see Section \ref{sec:eddington}).
}
\label{fig:CasSpec}
\end{figure}

The cascade emission $F(\epsilon)$ is calculated with Eq. \ref{eqn:cascade_full} in Appendix \ref{sec:cascade-form} following \citet{2011MNRAS.tmp..570T} and \citet{2011ApJ...733L..21D}.
For isotropy, the observed cascade emission has to be further modified with the solid angle
$\Omega_c \approx \pi\theta_c^2$ into which the intrinsic blazar emission is collimated where $\theta_c$ is the semi-aperture of the irradiated cone.
For blazars one has $\theta_c \sim 1/\Gamma_\mathrm{L}$ where $\Gamma_\mathrm{L}$ is the bulk Lorentz factor of the plasma of the jet. 
The observed emission is then found to be \citep{2011MNRAS.tmp..570T}
\begin{equation}
 F_\mathrm{obs}(\epsilon) = 2 \frac{\Omega_c}{4\pi}F(\epsilon),\label{eqn:CasEmi}
\end{equation}
where the factor of two accounts for the contribution of both jets in the isotropic case.
The exclusion criterion for an EBL shape at the 2$\sigma$ level reads
\begin{equation}
 F_\mathrm{obs}(\epsilon_\mathrm{meas}) > F_\mathrm{meas} + 2\sigma_\mathrm{meas},\label{eqn:cascade}
\end{equation}
where $\epsilon_\mathrm{meas}$, $F_\mathrm{meas}$, $\sigma_\mathrm{meas}$ are the measured energy, flux and statistical uncertainty reported in the 2FGL, respectively. 
In the case that the source is not detected $F_\mathrm{meas} = 0$ and $\sigma_\mathrm{meas}$ represents the $1\,\sigma$ upper limit on the flux. 
As an example, Figure \ref{fig:CasSpec} shows the observed and intrinsic VHE spectrum for a specific EBL shape of the blazar 1ES\,0229+200 together with the \emph{Fermi} upper limits \citep{2010MNRAS.406L..70T}.
The different model curves demonstrate the degeneracy between the different parameters entering the calculation. 
The EBL shape used to calculate the intrinsic VHE spectrum is not excluded in the isotropic case
since the emission does not overproduce the \emph{Fermi} upper limits. 
This is contrary to the case of $B_\mathrm{IGMF} = 10^{-20}$ G and $\Delta t = 3$ years where the predicted cascade flux exceeds the \fermilat~upper limits in the 1-10\,GeV range.
To obtain conservative upper limits of the EBL, only the isotropic case is assumed in the following, i.e. 
$B_\mathrm{IGMF} = 5 \times 10^{-13} \unit{G}$ which implies that the source has to be steady for a lifetime of $\Delta t \gtrsim 10^6$ years.
Furthermore, a Lorentz factor of $\Gamma_\mathrm{L} = 10$ is generically assumed for all sources.

\subsubsection{Total energy budget}
\label{sec:eddington}
The jets of AGN, the production sites for HE/VHE emission, are powered by the accretion of matter onto a central black hole \citep[e.g.][]{1995PASP..107..803U}.
If the radiation escapes isotropically from the black hole,
the balancing of the gravitational and radiation force leads to the maximum possible luminosity due to accretion, the Eddington luminosity, \citep[e.g.][]{2009herb.book.....D}
\begin{equation}
 L_\mathrm{edd}(M_\bullet) \approx 1.26 \times 10^{38} ~\frac{M_\bullet}{M_\odot} \unit{ergs}\unit{s}^{-1},
\end{equation}
where $M_\bullet$ is the black-hole mass normalized to the mass of the sun, $M_\odot$.
Assuming that the total emission of an AGN is not super-Eddington, the Eddington luminosity is the maximum power available for the two jets,
$P_\mathrm{jet} \le L_\mathrm{edd}/2$, which is a sum of several contributions which all can be represented as \citep[e.g.][]{1993MNRAS.264..228C,2011MNRAS.410..368B} 
 $P_i = \pi R^{\prime\,2}\Gamma^2 \beta c U^\prime_i$,
in the case that the radiation is emitted by an isotropically radiating relativistic plasma blob in the comoving frame.
The blob of 
radius $R^\prime$ in the comoving frame
moves with a bulk Lorentz factor $\Gamma_\mathrm{L}$ and corresponding speed $\beta c$, $U^\prime_i$ is the comoving energy density.
The energy density of the produced radiation is $U^\prime_\mathrm{r} = L^\prime/(4\pi R^{\prime\,2}c) = L / (4\pi \delta_\mathrm{D}^4R^{\prime\,2}c)$. 
The last equality connects the comoving luminosity with the luminosity in the lab frame via the 
Doppler factor $\delta_\mathrm{D} = [(1+z)\Gamma_\mathrm{L}(1-\beta\cos\theta)]^{-1} \approx 2\Gamma_\mathrm{L}/[(1+z)(1 + \theta^2\Gamma_\mathrm{L}^2)] $
where $\theta$ is the angle between the jet axis and the line of sight.
The approximation holds for $\theta \ll 1$ and $\Gamma_\mathrm{L} \gg 1$. 
Assuming $\theta \approx \theta_c$, the Doppler factor and the bulk Lorentz factor are equal up to the redshift factor $1 + z$, $\delta_\mathrm{D} \approx \Gamma_\mathrm{L}$.
The power produced in radiation is a robust lower limit for the entire power of the jet \citep[e.g.][]{2011MNRAS.410..368B}, 
\begin{equation}
P_\mathrm{r} \approx \fract{L}{(4\Gamma_\mathrm{L}^2)} < P_\mathrm{jet} \le L_\mathrm{edd}/2
\end{equation}
Solving the inequality for the observed luminosity, one arrives at an additional exclusion criterion for EBL shapes, 
namely, if the intrinsic energy flux at VHE is larger than the associated Eddington energy flux,
\begin{equation}
 (1 + z)^{2 - \Gamma_\mathrm{int} }\int\limits_{E_\mathrm{min}}^{E_\mathrm{max}}\Diff{N_\mathrm{int}}{E}\mathrm d E > \frac{\Gamma_\mathrm{L}^2L_\mathrm{edd}(M_\odot)}{2\pi d_\mathrm{L}^2}\label{eqn:eddington}
\end{equation}
where $E_\mathrm{min}$ and $E_\mathrm{max}$ are the minimum and maximum energy of the intrinsic VHE spectrum which is described with a power law with index $\Gamma_\mathrm{int}$.
The the factor $(1 + z)^{2-\Gamma_\mathrm{int}}$ accounts for the K-correction
and $d_\mathrm{L}$ is the luminosity distance given by
\begin{equation}
 d_\mathrm{L} = \frac{(1 + z)~c}{H_0} \int\limits_0^z \frac{\mathrm{d}z^\prime}{\sqrt{\Omega_m(1+z^\prime)^3 + \Omega_\Lambda}}.
\end{equation}
For a conservative estimate, $M_\bullet + \sigma_{M_\bullet}$ is used in the calculation of $L_\mathrm{edd}$.
The assumption of a non super-Eddington luminosity is, however, somewhat speculative as super-Eddington emission has been observed e.g. in the variable source 3C\,454.3 \citep{2011ApJ...733L..26A}.
In Section \ref{sec:results}, it will be shown that the capability of the Eddington criterion to exclude EBL shapes is extremely limited.
For this reason, only steady sources (listed in Table \ref{tab:steadysrc}) will be considered for this criterion.
Here, it is only emphasized that it is in principle possible to constrain the EBL with this argument.

Excluding EBL shapes with cascade emission (Eq. \ref{eqn:cascade}) and the total energy budget of the source (Eq. \ref{eqn:eddington}),
 will be referred to as the \textit{IntVHELumi} 
(short for intrinsic VHE luminosity) criterion.

\section{VHE AGN Sample}
\label{sec:samples}

\begin{table*}[thb]
\caption{VHE AGN spectra used in this study.
If not stated otherwise in the text, the \emph{Fermi} slope and variability index are taken from the 2FGL.}
\label{tab:samples}
\begin{scriptsize}
\begin{center}
\begin{tabular}{lcccccc|c|c}
\hline
\hline
\multirow{2}{*}{Source} & \multirow{2}{*}{Redshift} & \multirow{2}{*}{Experiment} & Energy Range & VHE Slope & \emph{Fermi} Slope & \multirow{2}{*}{Variability Index} & \multirow{2}{*}{Reference} & \multirow{2}{*}{Comments}\\
{} & {} & {} & (TeV) &  $\Gamma \pm \sigma_\mathrm{stat} \pm \sigma_\mathrm{sys}$ & $\Gamma \pm \sigma_\mathrm{stat}$& {} & {} & {}\\
\hline
Mkn\,421& 0.031 & HESS & 1.75 -- 23.1  & $2.05~\pm~0.22$ & $1.77~\pm~0.01$ & $112.8$ &  (1) & hardest index\\
Mkn\,501& 0.034 & MAGIC & 0.17 -- 4.43  & $2.79~\pm~0.12$ & $1.64~\pm~0.09$ & $72.33$ &  (2) & hardest index\\
Mkn\,501& 0.034 & HEGRA & 0.56 -- 21.45  & $1.92~\pm~0.03~\pm~0.20$ & $1.64~\pm~0.09$ & $72.33$  & (3) & hardest index\\
1ES\,2344+514& 0.044 & MAGIC & 0.19 -- 4.00  & $2.95~\pm~0.12~\pm~0.20$ & $1.72~\pm~0.08$ & $28.13$ &  (4) & steady \\
Mkn\,180& 0.045 & MAGIC & 0.18 -- 1.31  & $3.25~\pm~0.66$ & $1.74~\pm~0.08$ & $19.67$ &  (5) & steady \\
1ES\,1959+650& 0.048 & HEGRA & 1.52 -- 10.94  & $2.83~\pm~0.14~\pm~0.08$ & $1.94~\pm~0.03$ & $52.30$ &  (6) & hardest index\\
1ES\,1959+650& 0.048 & MAGIC & 0.19 -- 2.40  & $2.58~\pm~0.18$ & $1.94~\pm~0.03$ & $52.30$ &  (7) & hardest index\\
BL\,Lacertae& 0.069 & MAGIC & 0.16 -- 0.70  & $3.6~\pm~0.5$ & $2.11~\pm~0.04$ & $267.0$  & (8) & hardest index\\
PKS\,2005-489& 0.071 & HESS & 0.34 -- 4.57  & $3.20~\pm~0.16~\pm~0.10$ & $1.90~\pm~0.06$ & $68.86$  & (9) & hardest index\\
RGB\,J0152+017& 0.080 & HESS & 0.31 -- 2.95  & $2.95~\pm~0.36~\pm~0.20$ & $1.79~\pm~0.14$ & $27.73$ &  (10) & steady \\
PKS\,2155-304& 0.116 & HESS & 0.25 -- 3.20  & $3.34~\pm~0.05~\pm~0.1$ & $1.81~\pm~0.11$ & $262.9$  & (11) & simul \\
RGB\,J0710+591& 0.125 & VERITAS & 0.42 -- 3.65  & $2.69~\pm~0.26~\pm~0.20$ & $1.53~\pm~0.12$ & $29.86$  & (12) & steady \\
H\,1426+428& 0.129 & HEGRA & 0.78 -- 5.37  & -- & $1.32~\pm~0.12$ & $22.16$ &  (13) & steady\\
1ES\,0806+524& 0.138 & MAGIC & 0.31 -- 0.63  & $3.6~\pm~1.0~\pm~0.3$ & $1.94~\pm~0.06$ & $37.80$ &  (14) & steady \\
H\,2356-309& 0.165 & HESS & 0.23 -- 1.71  & $3.06~\pm~0.15~\pm~0.10$ & $1.89~\pm~0.17$ & $20.19$ &  (15) & steady \\
1ES\,1218+304& 0.182 & MAGIC & 0.09 -- 0.63  & $3.0~\pm~0.4$ & $1.71~\pm~0.07$ & $40.00$ &  (16) & steady \\
1ES\,1218+304& 0.182 & VERITAS & 0.19 -- 1.48  & $3.08~\pm~0.34~\pm~0.2$ & $1.71~\pm~0.07$ & $40.00$ &  (17) & steady \\
1ES\,1101-232& 0.186 & HESS & 0.18 -- 2.92  & $2.88~\pm~0.17$ & $1.80~\pm~0.21$ & $25.74$ &  (18) & steady \\
1ES\,1011+496& 0.212 & MAGIC & 0.15 -- 0.59  & $4.0~\pm~0.5$ & $1.72~\pm~0.04$ & $48.05$ &  (19) & hardest index\\
1ES\,0414+009& 0.287 & HESS & 0.17 -- 1.13  & $3.44~\pm~0.27~\pm~0.2$ & $1.98~\pm~0.16$ & $15.56$ &  (20) & steady \\
PKS\,1222+21~\tablefootmark{a}& 0.432 & MAGIC & 0.08 -- 0.35  & $3.75~\pm~0.27~\pm~0.2$ & $1.95~\pm~0.21$ & $13030$  & (21) & simul\\
3C\,279& 0.536 & MAGIC & 0.08 -- 0.48  & $4.1~\pm~0.7~\pm~0.2$ & $2.22~\pm~0.02$ & $2935$ &  (22) & hardest index\\
\hline
1ES\,0229+200~\tablefootmark{b}& 0.140 & HESS & 0.60 -- 11.45 & $2.5~\pm~0.19~\pm~0.10$ & -- & -- & (23) & -- \\
\hline
\end{tabular}
\tablefoot{See the text for details on the \textit{Comments} column.
\tablefoottext{a}{There was no simultaneous measurement during the 0.5\,h in which MAGIC detected the source. 
However, the index used in high energies was extracted from \emph{Fermi} data in the 2.5\,h before and after the MAGIC observation \citep{2011ApJ...730L...8A}.}
\tablefoottext{b}{The spectrum is only tested against the \intg criterion as it is not detected with the \fermilat.}
}
\tablebib{
(1)~{\citet{2011arXiv1106.1035T}};
(2)~{\citet{2011ApJ...727..129A}};
(3)~{\citet{1999A&A...349...11A}};
(4)~{\citet{2007ApJ...662..892A}};
(5)~{\citet{2006ApJ...648L.105A}};
(6)~{\citet{2003A&A...406L...9A}};
(7)~{\citet{2008ApJ...679.1029T}};
(8)~{\citet{2007ApJ...666L..17A}};
(9)~{\citet{2010A&A...511A..52H}};
(10)~{\citet{2008A&A...481L.103A}};
(11)~{\citet{2009ApJ...696L.150A}};
(12)~{\citet{2010ApJ...715L..49A}};
(13)~{\citet{2003A&A...403..523A}};
(14)~{\citet{2009ApJ...690L.126A}};
(15)~{\citet{2010A&A...516A..56H}};
(16)~{\citet{2006ApJ...642L.119A}};
(17)~{\citet{2009ApJ...695.1370A}};
(18)~{\citet{2006Natur.440.1018A}};
(19)~{\citet{2007ApJ...667L..21A}};
(20)~{\citet{2012arXiv1201.2044T}};
(21)~{\citet{2011ApJ...730L...8A}};
(22)~{\citet{2008Sci...320.1752M}};
(23)~{\citet{2007A&A...475L...9A}}
}
\end{center}
\end{scriptsize}
\end{table*}
In the past four years, the number of discovered VHE emitting AGN has doubled.
In this section samples of VHE spectra are defined that are evaluated with the \emph{VHE-HEIndex}, \pileup and \conca criteria (Section \ref{sec:sample-conc}) 
and with the \intg criterion (Section \ref{sec:sample-int}).

\subsection{Sample tested against concavity criteria}
\label{sec:sample-conc}
For this part of the analysis, 22 VHE spectra from 19 different sources are used. 
AGN are included in the sample only if their redshift is known, there is no confusion with other sources,
 and they are detected with the \fermilat. 
This excludes the known VHE sources 3C\,66A and 3C\,66B, 1ES\,0229+200, PG\,1553+113, and S\,50716+714.
Two spectra from the same source are only considered if they cover different energy ranges.
Furthermore, the radio galaxies Centaurus A and M~87 are not included since they are too close and measured at energies that are too low to yield any constraints of the EBL density.
Spectra that are a combination of several instruments are not included due to possible systematic uncertainties.
If two or more spectra are available for a variable source, the VHE spectrum is chosen that is 
measured simultaneously with \fermilat~observations. 
If the \emph{Fermi} spectrum is best described with a logarithmic parabola, the spectral index determined at the pivot energy is used for the comparison with the intrinsic VHE spectra. 
The entire AGN sample is listed in Table \ref{tab:samples} together with the redshift, the energy range, 
the spectral index at VHE energies, the index measured with the \fermilat, the variability index given in the 2FGL, and the corresponding references.

AGN are known to be variable sources both in overall flux and spectral index. 
This poses a problem for the \fermi criterion as it relies on the comparison of \fermilat~and IACT spectra. 
To address this issue, one can roughly divide the overall source sample into three categories: 

\begin{enumerate}
\item \emph{Steady sources in the \fermilat~energy band.}
In this category, all sources are assembled that show a variability index $<$ 41.64 in the 2FGL
which corresponds to a likelihood fit probability of more
 than 1\,\% that the source is steady \citep{2FGL}.
For these sources simultaneous measurements are not required
regardless if they are steady (like 1ES\,1101-232, \citealt{2007A&A...470..475A}) or variable (like H\,1426+428, see below) at VHE.
This does not affect the upper limits derived here because the \emph{Fermi} index remains valid as a lower limit independent of the VHE index.
These sources are marked as ``steady'' in the last column of Table \ref{tab:samples}.

\item \emph{Variable sources with simultaneous measurements.}
Some of the variable sources were observed simultaneously with the \fermilat~and IACTs in multiwavelength campaigns,
namely PKS\,2155-304 with HESS \citep{2009ApJ...696L.150A}, and PKS\,1222+21 with MAGIC \citep{2011ApJ...730L...8A}.
Instead of the spectral slopes given in the 2FGL, the \fermilat~spectra from these particular observations are used to test the EBL shapes.
These sources are marked as ``simul'' in the last column in Table \ref{tab:samples}.
Note, however, that the observation times might not be equal for the individual instruments since the sensitivities for the \fermilat~and IACTs are different.
Nevertheless, the arising systematic uncertainty is negligible for the sources under consideration.
In the case of PKS 2155-304, the source was observed in a quiescent state where no fast flux variability is expected.
PKS\,1222+21, on the other hand, was observed in a HE flaring state and \fermilat~observations are not available the 30 minutes of MAGIC observations.
Instead, \citet{2011ApJ...730L...8A} derive the \emph{Fermi} spectrum from 2.5\,hrs of data encompassing these 30 minutes.
This is justified, since the source remained in this high flux state for several days with little spectral variations \citep[cf. Figure 2 in][]{2011ApJ...733...19T}.
Accordingly, the maximum time lag allowed for observations to be considered as simultaneous is of the order of an hour.

\item \emph{Variable sources not simultaneously measured.}
For some variable sources, no simultaneous data are available, 
namely, 1ES\,1011+496, 1ES\,1959+650, 3C\,279, BL\,Lacertae, the flare spectra of Mkn\,501 and Mkn\,421,
and PKS\,2005-489 (see Table \ref{tab:samples} for the references). 
In these cases, the literature was examined for dedicated \fermilat~analyses of the corresponding sources in order to find hardest spectral index published. 
In the cases of 1ES\,1011+496, 1ES\,1959+650, PKS\,2005-489 and Mkn\,421 the indices reported in the 2FGL are the hardest published so far.
The hardest indices for BL\,Lacertae and Mkn\,501 are obtained by \citet{2010ApJ...716...30A} and \citet{2011ApJ...727..129A}, respectively, see Table \ref{tab:samples} for the corresponding values.
The distant quasar 3C\,279 was observed with the \fermilat~during a $\gamma$-ray flare in 2009 and the measured spectral indices
vary between $\sim 2$ and $\sim 2.5$ \citep[compare Fig. 1 in][]{2010Natur.463..919A}. 
Thus, the catalog index of $2.22\pm0.02$ is appropriate to use.
Table \ref{tab:samples} refers to all the spectra discussed here as ``hardest index'' in the last column.
\end{enumerate}
Additional uncertainties are introduced for the VHE observation with a maximum time lag between the measurement and the launch of the \emph{Fermi} satellite,
which is the case for Mkn\,501 and H\,1426+428.
In the case of H\,1426+428, no detection has been reported after the HEGRA measurement in 2002 at VHE which might suggest that the source is now in a quiescent state.
The 2002 spectrum with an observed spectral index of $\Gamma = 1.93 \pm 0.47$ is used in this study.
The hard spectrum promises stronger limits with the \fermi criterion than the 2000 spectrum which has a spectral slope of $\Gamma = 2.79 \pm 0.33$.
The source showed a change in flux by a factor of 2.5 between the 1999/2000 and 2002 observation runs but the spectral slope remained constant \citep{2003A&A...403..523A}.
Additionally, the \emph{Fermi} index of $1.32\pm0.12$ is the hardest of the entire sample and, in summary, it is chosen to include the source in the study.  
As for Mkn\,501, the spectrum of the major outbreak was measured up to 21\,TeV
and, consequently, it is a promising VHE spectrum to constrain the EBL density at FIR wavelengths.
As it turns out, it excludes most shapes due to the \pileup and \conca criteria.
These criteria are independent of the \emph{Fermi} index and, therefore, not affected by the difference in observation time. 

\subsection{Sample tested against intrinsic VHE luminosity}
\label{sec:sample-int}
For the integral criterion presented in Section \ref{sec:cascade} and \ref{sec:eddington},
only spectra from steady sources are used in order to avoid systematic uncertainties introduced by variability.
Only spectra are examined which suffer from large attenuations 
and are measured at energies beyond several TeV. 
These spectra are the most promising candidates for constraints 
as they show the highest values of integrated intrinsic emission. 
On the other hand, the spectrum of 1ES\,0229+200 which has been reanalyzed by \citet{2010MNRAS.406L..70T} can be tested against the \intg condition 
as upper limits on the HE flux suffice and no spectral information is required for this criterion. 
Otherwise, the same selection criteria apply as for the sample tested against concavity (known redshift, etc.).
The VHE spectra evaluated with the \intg criterion together with the central black-hole masses of the corresponding sources are summarized in Table \ref{tab:steadysrc}.

\begin{table}[tb]
\caption{Sources used to exclude EBL shapes with the \intg criterion.}
\label{tab:steadysrc}
\begin{center}
\begin{tabular}{lc}
\hline
\hline
\multirow{2}{*}{Source} &  Black-hole mass \\
{} & $\log_{10}(M_\bullet / M_\odot)$ \\
\hline
1ES\,0229+200  & $9.16~\pm~0.11$\\
1ES\,0414+009  & $9.3$\\
1ES\,1101-232  & $9$\\
1ES\,1218+304  & $8.04~\pm~0.24$\\
H\,1426+428    & $8.65~\pm~0.13$\\
H\,2356-309    & $8.08~\pm~0.23$\\
RGB\,J0152+017 & $9$\\
RGB\,J0710+591 & $8.25~\pm~0.22$\\
\hline
\end{tabular}
\end{center}
\tablefoot{The black hole masses $M_\bullet$ are taken from \citet{2008MNRAS.385..119W} except from RGB\,J0710+591
and 1ES\,0414+009  for which the masses are given in \citet{2005ApJ...631..762W} and \citet{2000ApJ...532..816U}, respectively.
No measurements of the central black hole masses of 1ES\,1101-232 and RGB\,J0152+017 are available so the fiducial value of $M_\bullet = 10^9M_\odot$ is used here.
}
\end{table}
\begin{figure*}[htb]
 \centering
 \includegraphics[width = .33\linewidth , angle=270]{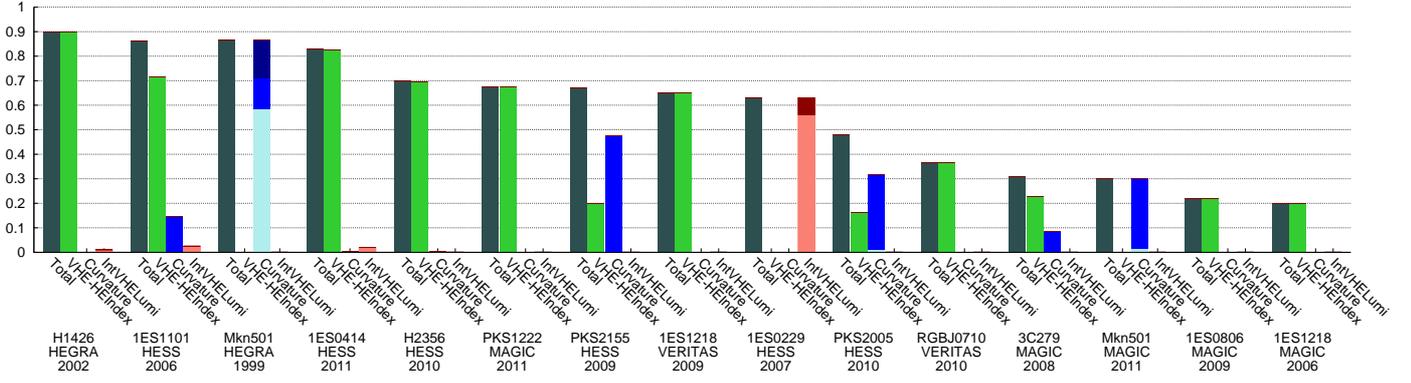}
\caption{Histogram of the fraction of excluded shapes of the different VHE spectra. 
The columns show the total fraction of rejected shapes as well as the fraction excluded by the different criteria.
The column labeled ``Curvature'' combines the \conca and \pileup criteria.
Spectra that allow more than 90\,\% of all shapes are not shown.}
\label{fig:Hist}
\end{figure*}

\section{Results}
\label{sec:results}
The upper limits on the EBL density are derived by calculating the envelope shape of all \textit{allowed} EBL shapes. 
The influence of the different exclusion criteria is examined by inspecting the envelope shape due to the \fermi argument alone and successively adding the other criteria and reevaluating the resulting upper limits. 
Furthermore, the impact of the VHE spectra responsible for the most stringent limits in the optical, MIR and FIR will be investigated by excluding these spectra from the sample and inspect the change in the upper limits.

Figure \ref{fig:Hist} shows a histogram of the fractions of rejected shapes by each VHE spectrum, 
where the different colors represent the different criteria that lead to the exclusion of an EBL shape.
It should be noted that individual shapes can be rejected by several criteria at the same time, and, therefore, the different columns may add up to a number larger than indicated by the total column.
Results for spectra that exclude no (BL\,Lacertae, 1ES\,2344+514, and Mkn\,180) or less than 10\,\% of all EBL shapes 
(the MAGIC spectrum of 1ES\,1959+650, the HESS spectra of RGB\,J0152+017 and Mkn 421, as well as the HEGRA spectrum of 1ES\,1959+650)
are not shown.
Most EBL shapes are excluded by the VHE spectra of 
H\,1426+428, 1ES\,1101-232, and Mkn\,501. 
The influence of H\,1426+428, and Mkn\,501 on the limits in the MIR and FIR and of 3C\,279 together with PKS\,1222+21 in the optical will be examined by excluding these spectra from the sample.
These sources provide strong constraints in the respective wavelength bands.
Note that removing  1ES\,1101-232 from the source sample does not change the upper limits 
since a number of spectra of sources with comparable redshifts (cf. Table \ref{tab:samples}) exclude the same EBL shapes as 1ES\,1101-232, e.g. 
1ES\,0414+009, the VERITAS spectrum of 1ES\,1218+304, H\,2356-309, and PKS\,2005-489.

\begin{figure*}[htb]
 \centering
 \includegraphics[width = .7 \linewidth, angle = 270]{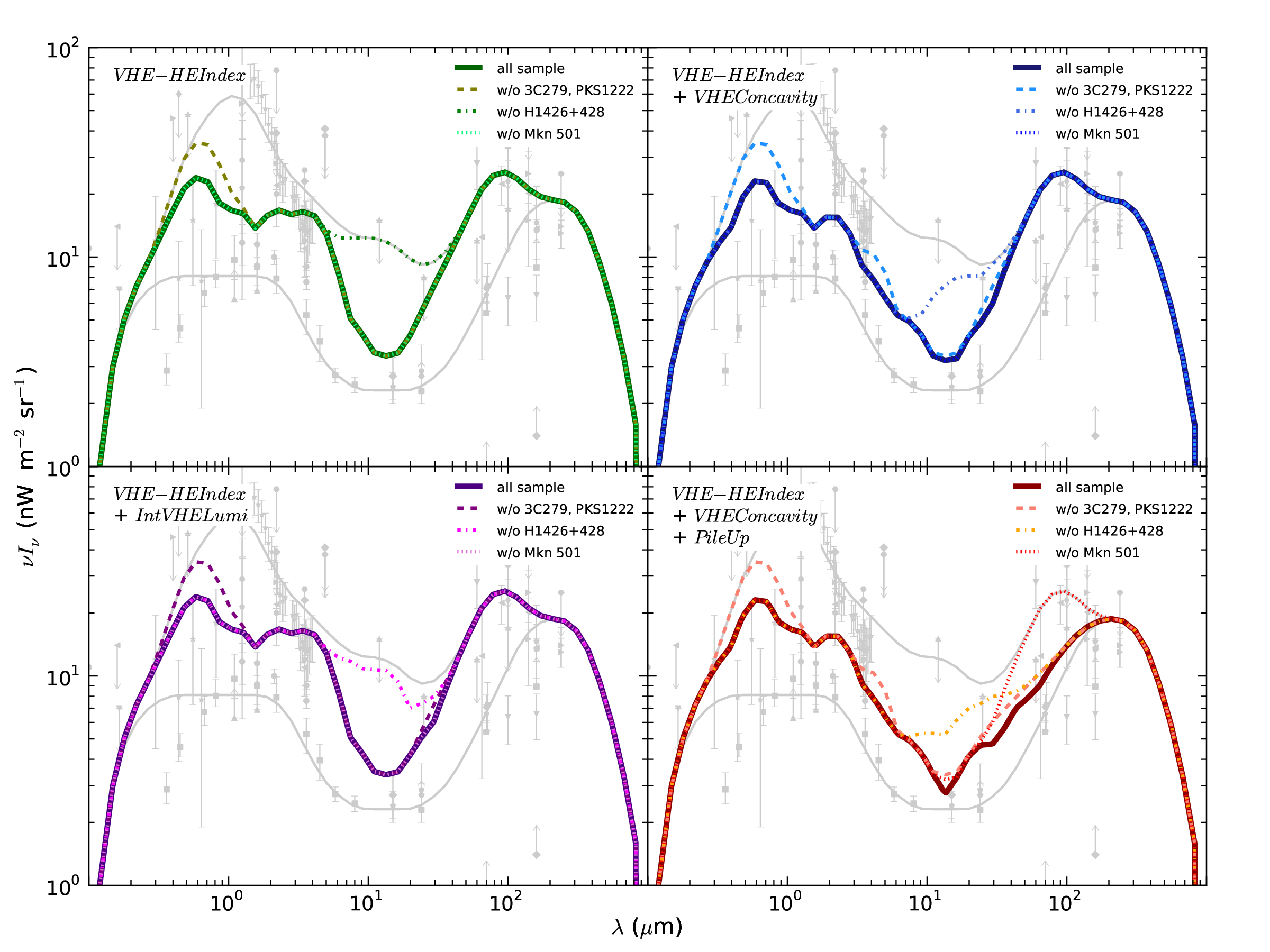}
\caption{Limits on the EBL density for different exclusion criteria.
The solid line is the envelope shape of all allowed shapes from the combination of all VHE spectra whereas the dashed curve shows the envelope shape without considering 
the VHE spectrum of 3C\,279. The dashed dotted line is the envelope shape without H\,1426+428 and the dotted line displays the upper limits without Mkn\,501.}
\label{fig:CompCriteria}
\end{figure*}
Different combinations of exclusion criteria are shown in the panels of Figure \ref{fig:CompCriteria}. 
Each panel depicts the limits for the complete spectrum sample and, additionally, the resulting EBL constraints if the spectra discussed above are omitted.
By itself, the \fermi criterion gives strong upper limits in the optical and MIR on the EBL density if all spectra are included (upper left panel of Figure \ref{fig:CompCriteria}).
In the optical, the limits are dominated by the spectra of 3C\,279 and PKS\,1222+21, so,
consequently, the restrictions are significantly weaker without these spectra (dashed line in Figure \ref{fig:CompCriteria}).
The spectra are influenced most by changes of the EBL density in the optical which is inferred from the maximum energies of 480\,GeV and 350\,GeV for 3C\,279 and PKS\,1222+21, respectively.
They translate into maximum cross sections for pair production at 0.6\,\murm m (3C\,279) and 0.43\,\murm m (PKS\,1222+21), see Eq. \ref{eqn:ppwave}.
The constraints are almost unaltered if only one of these spectra is excluded from the sample.
In the MIR, the spectrum of H\,1426+428 provides firm limits on the EBL density whereas
scarcely any EBL shape is rejected due to the spectrum of Mkn\,501 with the \fermi criterion.

\begin{figure}[tb]
 \centering
 \includegraphics[width = 0.915\linewidth]{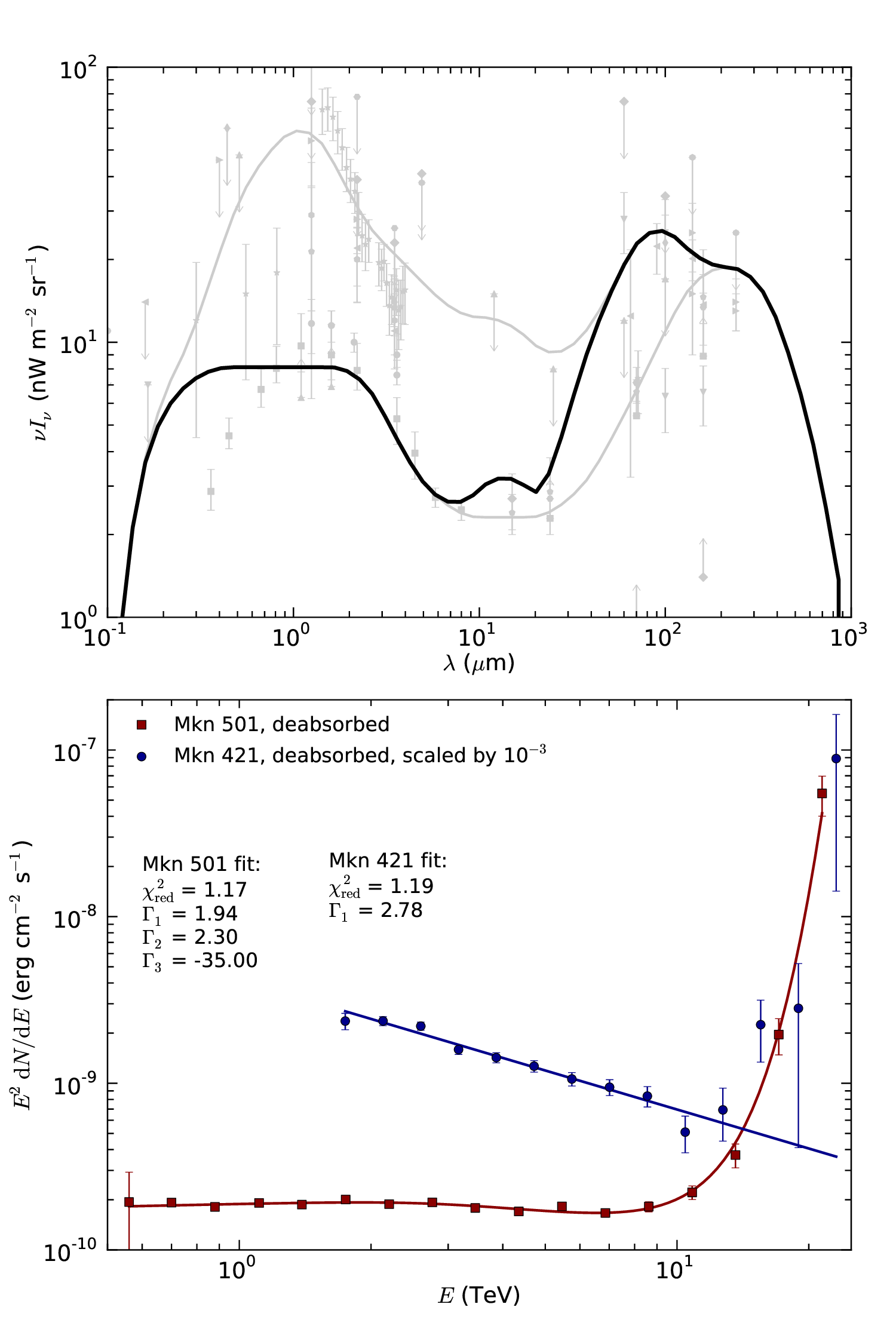}
\caption{Upper panel: Example of an EBL shape excluded by Mkn\,501 with the \conca criterion. 
Lower panel: The spectra of Mkn\,501 and Mkn\,421 corrected with this particular EBL shape. The flux of the latter is scaled by $10^{-3}$ for better visibility.
For Mkn\,501, a double broken power law provides the best description with a spectral index $\Gamma_3=-35$ at highest energies, the maximum value tested in the fitting procedure. 
In the case of Mkn\,421, a simple power law suffices.}
\label{fig:Mkn501}
\end{figure}

The combination of the \fermi and \conca criterion strengthens the upper limits between 2\,\murm m and 10\,\murm m,
as shown in the upper right panel of Figure \ref{fig:CompCriteria}.
Convex intrinsic spectra are the result of an EBL density with a positive gradient between lower and higher wavelengths and, thus, 
a combination with the \fermi criterion is necessary to exclude shapes with a high EBL density that are rather constant in wavelength.
Therefore, on their own, neither the \pileup nor the \conca criterion provide strong upper limits.
Combining the \pileup and \fermi arguments results in very similar limits as the combination of the \fermi and \conca criterion.
This degeneracy between the \pileup and \conca criterion is also demonstrated in Figure \ref{fig:Mkn501}.
The spectrum of Mkn\,501 corrected with a certain EBL shape shows a strong exponential rise at highest energies but is best described with a double broken power law. 
The combination of the \pileup and \conca together with the \fermi criterion yields robust upper limits in the FIR
as displayed in the lower right panel of Figure \ref{fig:CompCriteria}.
The constraints in the FIR are entirely due to the spectrum of Mkn\,501
although the spectrum of Mkn\,421 is also measured beyond 20\,TeV and both sources have a comparable redshift.
However, the spectrum of Mkn\,421 rejects far less shapes than Mkn\,501.
Indeed, an exponential rise is observed in intrinsic spectra of Mkn\,421 for certain EBL realizations (e.g. the corrected Mkn\,421 spectrum in Figure \ref{fig:Mkn501})
but a power law is found to be the best description of the spectrum.

\begin{figure}[tb]
 \centering
 \includegraphics[width = .75 \linewidth, angle = 270]{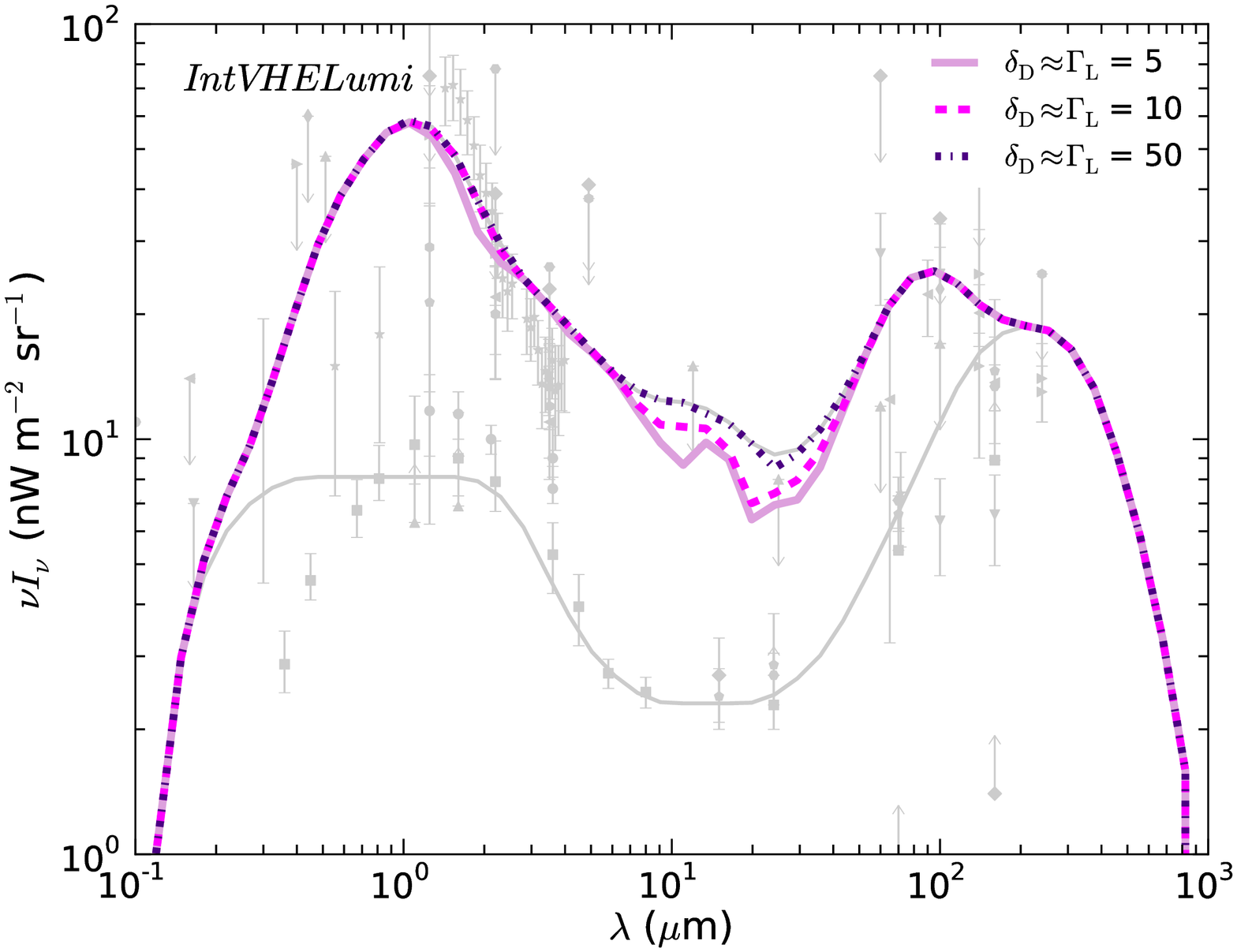}
\caption{Upper limits solely due to the \intg criterion for different values of the bulk Lorentz factor $\Gamma_\mathrm{L}$ and the Doppler factor $\delta_\mathrm{D}$ of the emitting region.
With increasing Lorentz and Doppler factor, respectively, the limits become worse (see Eqs \ref{eqn:CasEmi} and \ref{eqn:eddington}).
}
\label{fig:CompIntegral}
\end{figure}

Compared to the \fermi criterion alone, 
the combination with the \intg criterion leads to improved upper limits 
only if H\,1426+428 is discarded from the sample 
(lower left panel of Figure \ref{fig:CompCriteria}).
Most shapes are rejected by the VHE spectrum of 1ES\,0229+200 which is also the sole spectrum which excludes a very limited number of shapes with the Eddington luminosity argument.
Remarkably, the spectrum 
excludes 
more than 60\,\% of all shapes.
The \intg criterion has the most substantial effect in the infrared part of the EBL density as the highest energies of the spectrum of 1ES\,0229+200 contribute most to the integral flux.
The maximum energy measured in the spectrum is 11.45\,TeV and thus the limits are most sensitive to changes in the EBL around 14\,\murm m.
The influence of the choice of the bulk Lorentz factor $\Gamma_\mathrm{L}$ (and hence of the Doppler factor $\delta_\mathrm{D}$ since $\Gamma_\mathrm{L} \approx \delta_\mathrm{D}$ is assumed)
on the envelope shape can be seen from Figure \ref{fig:CompIntegral}, where the upper limits are shown for $\Gamma_\mathrm{L} =$ 5, 10, and 50. 
As $\Gamma_\mathrm{L}$ enters quadratically into the calculation of the flux (cf. Eq. \ref{eqn:CasEmi}) 
and for the Eddington luminosity (Eq. \ref{eqn:eddington})
the choice of the value of $\Gamma_\mathrm{L}$ is critical for the number of rejected EBL shapes.
The bulk Lorentz factor is unknown for the sources tested with \intg~and for the combination with the other criteria $\Gamma_\mathrm{L} = 10$ is generically chosen.
However, even with this oversimplified choice of $\Gamma_\mathrm{L}$, the \intg~criterion does not lead to improvements of the upper limits compared to the combination of the \fermi, \conca~and \pileup~criteria.
Conversely, this implies that the final upper limits will not depend on the specific choice of model parameters and assumptions that enter the evaluation of the \intg~criterion.

\begin{figure*}[htb]
 \centering
 \includegraphics[width = .75 \linewidth, angle = 270]{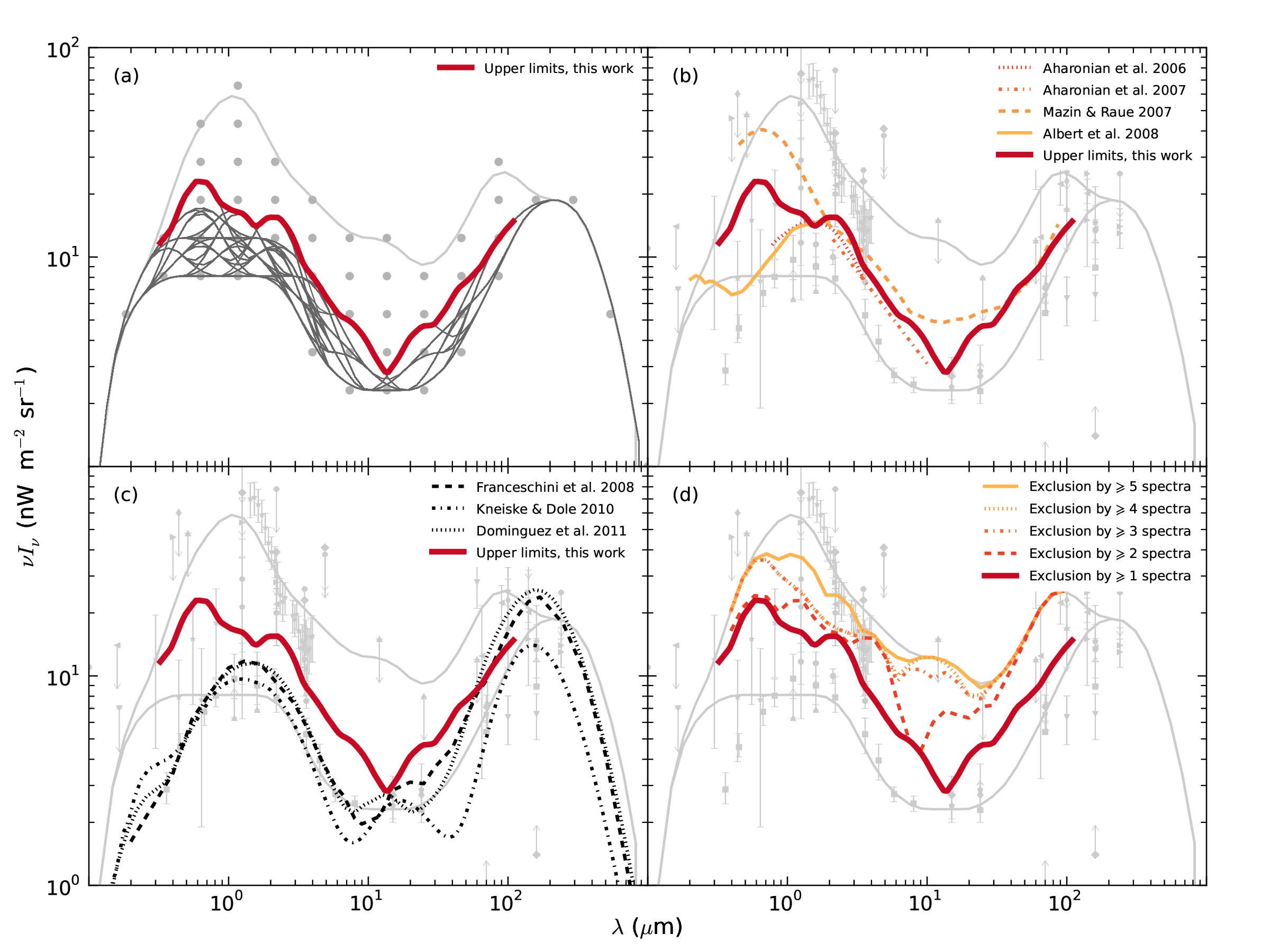}
\caption{
Upper limits derived in this study. 
(a) The envelope shape (upper limits) of all allowed shapes (dark gray lines). Also shown are the grid points as light gray bullets.
(b) The constraints compared to the the upper limits of MR07, \citet{2006Natur.440.1018A,2007A&A...475L...9A}, and \citet{2008Sci...320.1752M}.
(c) Upper limits of this study together with three EBL models \citep{2008A&A...487..837F,2010A&A...515A..19K,2011MNRAS.410.2556D}.
(d) Upper limits requiring different minimum numbers of VHE spectra that exclude an EBL shape.
}
\label{fig:master-result}
\end{figure*}

The final result for the upper limits is the combination of all criteria and all VHE spectra, shown in Figure \ref{fig:master-result}.
It is the envelope shape of all allowed EBL realizations, cf. Figure \ref{fig:master-result}a,
which itself is excluded by several VHE spectra and it should, thus, not be regarded as a possible level of the EBL density.
For the maximum energy of all VHE spectra of 23.1\,TeV, the cross section for pair production peaks at a wavelength of the EBL photons of $\lambda_\ast\approx 29$\,\murm m. 
More than half of the interactions occur in a narrow interval $\Delta\lambda = (1 \pm 1/2)\lambda_\ast$ around the peak wavelength \citep[e.g][]{2006Natur.440.1018A} 
and hence the constraints are not extended beyond 100\,\murm m.
Albeit including the evolution of the EBL with redshift, the derived upper limits are below 5\,nW\,m$^{-2}$\,sr$^{-1}$ in the range from 8\,\murm m to 31\,\murm m.
A comparison of the constraints with previous works is shown in Figure \ref{fig:master-result}b.
Above 30\,\murm m, the constraints are consistent with those derived in MR07. 
For wavelengths between 1\,\murm m and 4\,\murm m the limits are in accordance with the results of \citet{2006Natur.440.1018A,2007A&A...475L...9A}
and \citet{2008Sci...320.1752M}. 
The strong limits of \citet{2008Sci...320.1752M}
 who utilized the spectrum of 3C\,279
are not reproduced. Note, however, that these limits are derived by changing 
certain free parameters (e.g. fraction of UV emission escaping the galaxies) of the EBL model of \citet{2002A&A...386....1K} while the current approach allows for generic EBL shapes.
Consequently, an EBL shape with a high density at UV / optical wavelengths 
followed by a steep decline towards optical and NIR wavelengths produces a soft intrinsic spectrum of 3C\,279 that cannot be excluded by any criterion.
Furthermore, an inspection of the spectrum of 3C\,279 shows that the fit will be dominated by the first two energy bins due to the smaller error bars.
Thus, a convex spectrum is often still sufficiently described with a soft power law.
In general, it should be underlined that all of the above limits from recent studies use a theoretically motivated bound on the intrinsic spectral slope of $\Gamma = 1.5$.

\begin{figure*}[htb]
 \centering
 \includegraphics[width = .6 \linewidth, angle = 270]{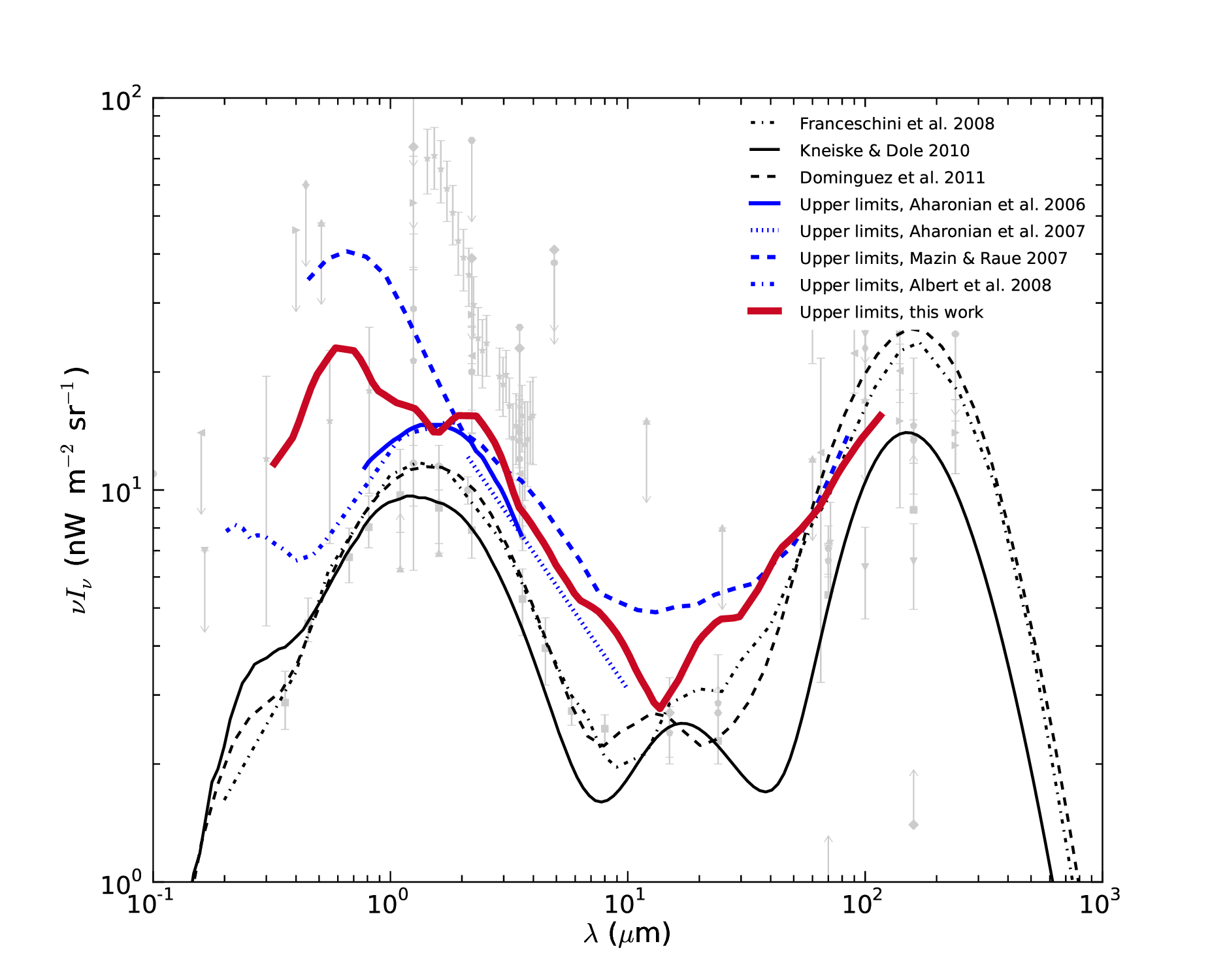}
\caption{
Upper limits of this work together with previous limits and EBL models.
}
\label{fig:master-result-app}
\end{figure*}

The upper limits derived here are not in conflict with the EBL model calculations of \citet{2008A&A...487..837F}, \citet{2010A&A...515A..19K} and \citet{2011MNRAS.410.2556D} 
and are compatible with the lower limit galaxy number counts derived from Spitzer measurements \citep{2004ApJS..154...39F} (see Figure \ref{fig:master-result}c).
In the FIR, the models of \citet{2008A&A...487..837F} and \citet{2011MNRAS.410.2556D} lie above the derived upper limits, though one should note that the EBL limit at these wavelengths relies on a single spectrum (Mkn\,501).
Between $\sim$\,1\,\murm m and $\sim$\,14\,\murm m there is, however, a convergence between the upper limits and model calculations and at 13.4\,\murm m
the EBL is constrained below 2.7\,nW\,m$^{-2}$\,sr$^{-1}$, just above the EBL models.
This leaves not much room for additional components such as Population III stars
and implies that the direct measurements of \citet{2005ApJ...626...31M} are foreground dominated as discussed in \citet{2005ApJ...635..784D}. 
The EBL models, the upper limits from previous works, and the results derived here are shown together in Figure \ref{fig:master-result-app}.

\begin{figure}[htb]
 \centering
 \includegraphics[width = 0.65 \linewidth,angle = 270 ]{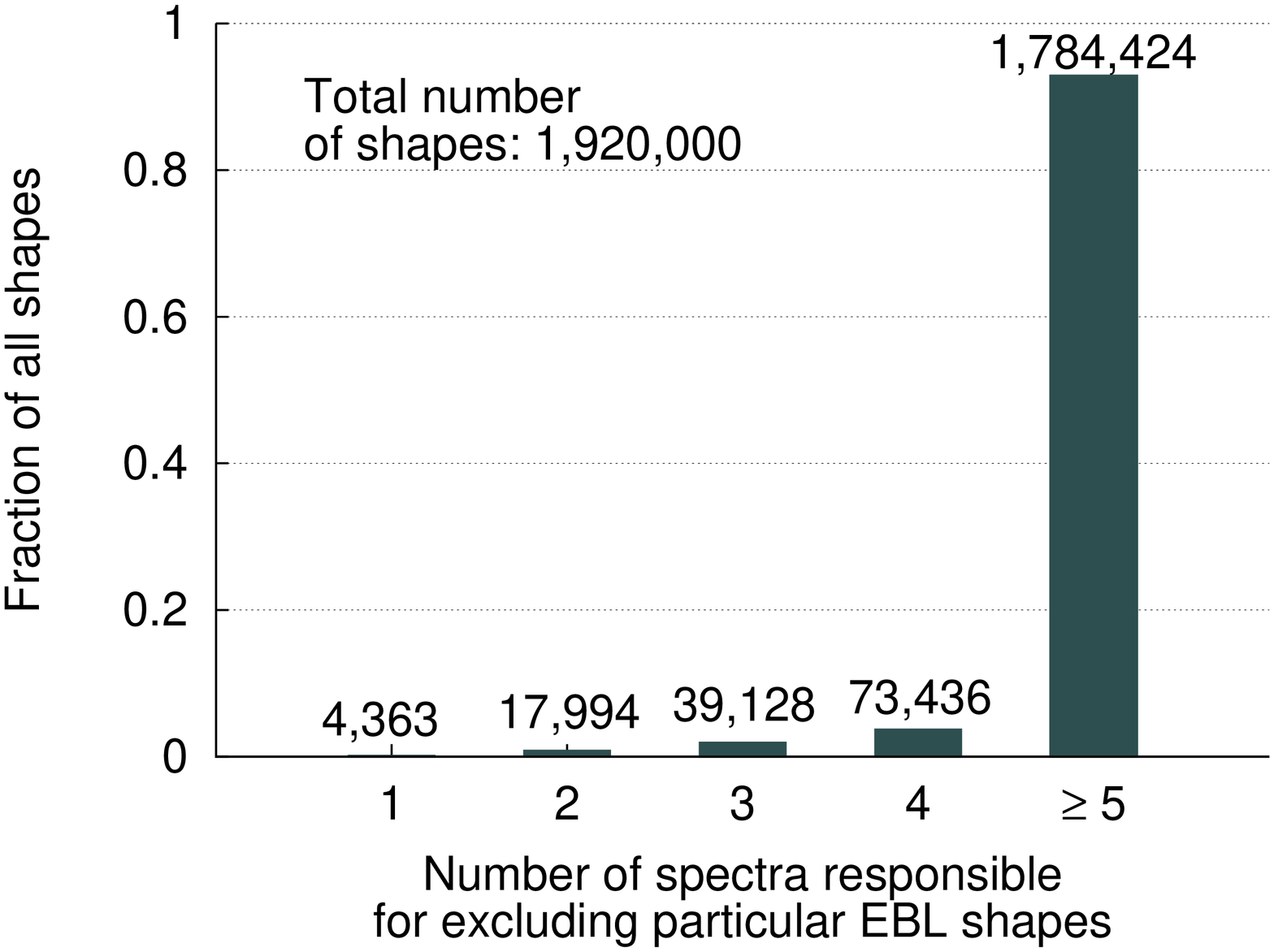}
\caption{
Percentage of all shapes excluded by at least a certain number of spectra. The majority of rejected shapes is not allowed by more than five spectra.
}
\label{fig:counthist}
\end{figure}

In general, most of the tested EBL shapes are excluded by more than one spectrum (Figure \ref{fig:counthist}).
While 0.23\,\% of all EBL shapes are excluded by only one of the spectra in the sample, the majority of shapes (93\,\%) is rejected by five spectra or more.
Figure \ref{fig:master-result}d shows the limits for different minimum numbers of VHE spectra that rule out an EBL shape.
From NIR to MIR wavelengths, the limits are only slightly worsened 
if at least two spectra are required to exclude EBL shapes.
If at least five spectra are ought to reject an EBL shape,
the EBL density remains confined below 40 nW\,sr$^{-1}$\,m$^{-2}$ in the optical. 
Thus, from optical to MIR wavelengths, the limits are robust against individual spectra that possibly have a peculiar intrinsic shape
due to one of the mechanisms discussed in Section \ref{sec:curvature}.
Especially in the MIR and FIR, however, the limits are weakened as they mainly depend on two spectra, H\,1426+428 and Mkn\,501.
This underlines the need for more spectra measured beyond several TeV in order to draw conclusions 
about the EBL density in the MIR and FIR from VHE blazar measurements.

Given the similarities in procedures used, 
the systematic uncertainties of the limits derived in this study are similar to the ones derived in MR07 and \citet{}.
They have been estimated to be 31\,\% in optical to near infrared and 32 - 55\% in mid to far infrared wavelengths,
mainly from the grid spacing and the uncertainties on the absolute energy scale of ground based VHE instruments which is taken to be 15\,\%.
Note that \citet{2010A&A...523A...2M} achieved a cross-calibration using the broadband SED of the Crab Nebula between the \fermilat~and IACTs by shifting the IACT measurements by $\sim$\,5\,\% in energy. 
As it turns out, 
additional uncertainties arise from the phenomenological description of the EBL evolution 
($<$ 4\,\% for a redshift $z = 0.2$ and $<$ 10\,\% for $z = 0.5$, \citealt{2008IJMPD..17.1515R}).
Uncertainties in the calculation of the cascade emission are caused by the choice of the model parameters which are, however, difficult to quantify.
The same applies for the assumption that steady sources do not show super-Eddington luminosities.
Since the most stringent limits do not rely on these exclusion criteria, these uncertainties do not affect the final results of the upper limits.

Additionally, the measurement capabilities of the \fermilat~affect the \fermi~criterion and hence the upper limits.
While the 2FGL does not quote the systematic errors on the individual spectral indices, it gives a number of sources of systematic errors:
the effective area, the diffuse emission model, and the handling of front and back converted events.
The systematic error on the effective area is estimated to be between 5\,\% and 10\,\%, while the errors on the diffuse emission model mainly effects sources inside the galactic plane.
Furthermore, the isotropic emission for front and back converted events is assumed to be equal. 
This leads to underestimation of the flux below 400\,MeV and might produce harder source spectra. 
As harder spectra in the \fermilat~band weaken the upper limits, the results derived here can, again, be regarded as conservative.

\section{Summary and Conclusions}
\label{sec:concl}

In this paper, new upper limits on EBL density over a wide wavelength range from the optical to the far infrared are derived,
utilizing the EBL attenuation of HE and VHE \grays~from distant AGN.
A large number of possible EBL realization is investigated, allowing for possible features from, e.g., the first stars.
Evolution of the EBL density with redshift is taken into account in the calculations using a phenomenological prescription \citep[see e.g.][]{2008IJMPD..17.1515R}.
A large sample of VHE spectra consisting of 23 spectra from 20 different sources
with redshifts ranging from $z = 0.031$ to $0.536$ is used in the analysis.
The VHE spectra are corrected for absorption and subsequently investigated for their physical feasibility.
Two basic criteria are examined: (1) concavity of the high energy part of the spectrum spanning from HE to VHE and 
(2) total integral flux in the VHE, a novel way to probe the EBL density.
For the former criterion, spectra from the 
\fermilat~at HE are used as a conservative upper limit, combined with criteria on the overall VHE concavity.
This is a more conservative argument than a theoretically motivated bound on the intrinsic spectral index at VHE of, say, $\Gamma = 1.5$.
This value, used in previous studies, is somewhat under debate as a harder index can be possible, for instance, 
if the underlying population of relativistic electrons is very narrow \citep{2006MNRAS.368L..52K,2009MNRAS.399L..59T,2011arXiv1106.4201L},
in the case of internal photon absorption \citep{2008MNRAS.387.1206A}, or in proton-synchrotron models \citep[e.g.][]{2000NewA....5..377A,2011ApJ...738..157Z}.
For the latter criterion, the expected cascade emission is investigated and, 
additionally, the total intrinsic luminosity is compared to the Eddington luminosity of the AGN.
Limits on the EBL density are derived using each of the criteria individually and for combinations of the criteria. 
In addition, the influence of individual data sets is tested.
The obtained constraints reach from 0.4\,\murm m to 100\,\murm m and are
below 5\,nW\,m$^{-2}$\,sr$^{-1}$ between 8\,\murm m and 31\,\murm m
even though more conservative criteria are applied and the evolution of the EBL with redshift is accounted for.
In the optical, the EBL density is limited below 24 nW\,m$^{-2}$\,sr$^{-1}$.

The limits forecast
a low level of the EBL density from near to far infrared wavelengths also predicted by the models of \citet{2010A&A...515A..19K} and \citet{2011MNRAS.410.2556D}
which is in accordance with MR07.
Furthermore, the constraints exclude the direct measurements of \citet{2005ApJ...626...31M}. 
Certain mechanisms, however, are discussed in the literature that effectively reduce the the attenuation of \grays~due to pair production. 
For instance, if cosmic rays produced in AGN are not deflected strongly in the intergalactic magnetic field
they could interact with the EBL and form VHE \grays~that contribute to the VHE spectrum \citep{2010APh....33...81E,2010PhRvL.104n1102E,2011ApJ...731...51E}.
Other suggestions are more exotic as they invoke the conversion of photons into axion like particles \citep[e.g][]{2009MNRAS.394L..21D,2009JCAP...12..004M} 
or the violation of Lorentz invariance \citep[e.g.][]{2008PhRvD..78l4010J}.

Future simultaneous observations of extragalactic blazars with the \fermilat~and IACTs have the potential to further constrain the EBL density.
\begin{acknowledgements}
MM would like to thank the state excellence cluster ``Connecting Particles with the Cosmos'' at the university of Hamburg.
The authors would also like to thank the anonymous referee for helpful comments improving the manuscript.
\end{acknowledgements}

\appendix
\section{Evolution of the EBL with redshift}
\label{sec:evo}
\begin{figure}[tb]
 \centering
 \includegraphics[width = .75 \linewidth, angle = 270]{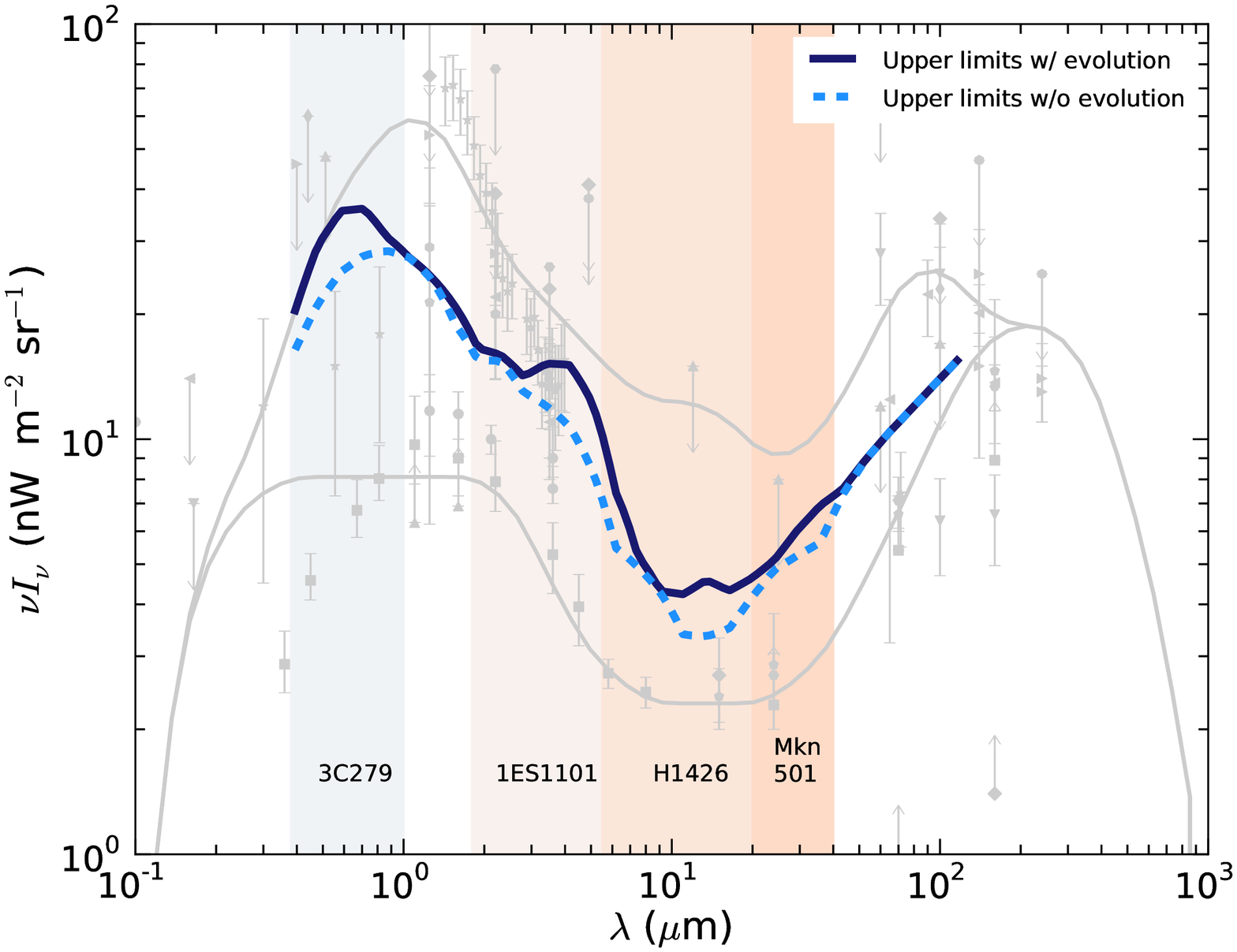}
\caption{
Upper limits with and without evolution for four VHE spectra. The individual spectra have the strongest influence in the correspondingly shaded regions.
}
\label{fig:evo-noevo}
\end{figure}

In this Appendix the influence of the evolution with redshift of the EBL density on the upper limits is investigated.
For this purpose, the limits are calculated for four VHE spectra, 
namely 1ES\,1101-232, H\,1426+428, Mkn\,501, 
and 3C\,279 which is the source with the largest redshift of $z = 0.536$ in the sample.
The evolution is included with a phenomenological ansatz \citep[e.g.][]{2008IJMPD..17.1515R}.
The four spectra are tested against the \fermi, \pileup and \conca criteria 
and the envelope shape is determined with and without the evolution with redshift.
Not surprisingly, the EBL density is less confined if the evolution is accounted for, as seen from Figure \ref{fig:evo-noevo}.
The differences are most pronounced in the optical, where the influence of 3C\,279 is the strongest (light blue shaded region in Figure \ref{fig:evo-noevo}). 
Without taking the evolution into account, the limits are overestimated by up to 40\,\% at 0.6\,\murm m. 
At higher wavelengths, the difference is not as distinct as in the optical. 
This outcome emphasizes that the evolution of the EBL density with redshift has a non-negligible effect on upper limits derived from VHE AGN spectra,
especially for sources with a large redshift. 

\section{Cascade emission formulae}
\label{sec:cascade-form}
The electron spectrum due to the interaction of VHE photons with the EBL is obtained by solving the corresponding kinetic equation for the electron distribution $N(\gamma)$
in the steady state limit
\citep{2011MNRAS.tmp..570T},
\begin{equation}
 N(\gamma) = \frac{1}{|\dot\gamma|}\int\limits_\gamma^\infty\mathrm d \gamma^\prime Q(\gamma^\prime)
\end{equation}
where $|\dot\gamma|$ is the energy loss due to IC scattering with CMB photons in the Thomson regime with an injection rate
\begin{equation}
 Q(\gamma) = \Diff{N_\mathrm{obs}}{E}\left(e^{\tau_\gamma(E,z)} - 1\right). 
\end{equation}
The approximation $E = 2m_ec^2\gamma$ is used, with $m_e$ the mass of the electron.
The cascade spectrum for scattered photons with energy $\epsilon$ off the $e^+e^-$ pairs is then readily calculated to be \citep{1970RvMP...42..237B}
\begin{eqnarray}
 F(\epsilon) &=&\frac{9}{64} \frac{\epsilon~m_ec^2}{u_{\mathrm{CMB}}(z = 0)} \nonumber \\
&{}&\times\int\limits_{\mathrm{max}\left[\sqrt{\epsilon / 4\epsilon_\mathrm{CMB}},\gamma_\mathrm{dfl},
\gamma_\mathrm{eng}\right]}^\infty \hskip-10pt \frac{\mathrm d\gamma}{\gamma^6}
\int\limits_{2m_ec^2\gamma}^\infty \mathrm d E \Diff{N_\mathrm{obs}}{E}\left(e^{\tau_\gamma(E,z)} - 1\right) \nonumber \\
&{}&\times\int\limits_0^\infty \mathrm d \epsilon^\prime \frac{n_\mathrm{CMB}(\epsilon^\prime, z = 0)}{\epsilon^{\prime\,2}} F_\mathrm{T}^\mathrm{IC}(\epsilon,\epsilon^\prime,\gamma),
\label{eqn:cascade_full}
\end{eqnarray}
where $u_\mathrm{CMB}(z=0) = 0.26\unit{eV}\unit{cm}^{-3}$, $\epsilon_\mathrm{CMB} (z=0)= 634$\,\murm eV and $n_\mathrm{CMB}(\epsilon^\prime,z)$ denote the energy density, 
mean energy and differential photon number density of the CMB, respectively.
The inverse Compton kernel for scattering on an isotropic photon field in the Thomson regime 
(Klein-Nishina effects can be neglected for the energies of the primary photons considered here) is
\begin{equation}
 F_\mathrm{T}^\mathrm{IC}(\epsilon,\epsilon^\prime,\gamma) = 4\epsilon^\prime\gamma\left(2\hat{\epsilon}\ln\hat\epsilon + \hat\epsilon + 1 -2\hat\epsilon\right),
\end{equation}
with $0 \le F_\mathrm{T}^\mathrm{IC} \le 1$ and $\hat\epsilon = \epsilon / (4\epsilon^\prime\gamma^2)$. The lower limit for the integration over $\gamma$ is the maximum 
of three different constraints on $\gamma$. The first one stems from kinematic constraints of Compton scattering. The second denotes the $\gamma$-factor for which the 
electrons are deflected outside the opening cone of the blazar jet with an opening angle $\theta_c \sim 1/\Gamma_\mathrm{L}$, with $\Gamma_\mathrm{L}$ the bulk Lorentz factor of the plasma of the jet. 
And finally, the third factor gives the minimum possible Lorentz factor if the source is active for a certain time \citep[see][for further details]{2011ApJ...733L..21D}.
However, $\gamma_\mathrm{dfl}$ and $\gamma_\mathrm{eng}$ are calculated by \citet{2011ApJ...733L..21D} under the approximation of small deflection and observing angles.
In the case of isotropic emission, i.e. $B_\mathrm{IGMF} \approx 10^{-13}$ G and a lifetime of the source of $\Delta t \gtrsim 10^{6}$ years, 
this approximation does not longer hold and 
 the lower integration bound is replaced by $\sqrt{\epsilon / 4\epsilon_\mathrm{CMB}}$.

\bibliographystyle{aa}
\bibliography{vhe_spectra,aa00000}

\end{document}